\documentclass[12pt]{article}

\usepackage[letterpaper,margin=1.25in]{geometry}

\usepackage{authblk}

\usepackage[parfill]{parskip}
\usepackage{setspace}

\usepackage[labelfont=bf,font={small,stretch=0.95}]{caption}

\usepackage[super,sort,compress,numbers]{natbib}

\usepackage{hyperref}
 
\usepackage{textcomp, gensymb}
\usepackage{amsmath}
\usepackage{amssymb}

\usepackage{graphicx}
\usepackage{multirow}
\PassOptionsToPackage{hyphens}{url}\usepackage{hyperref}

\usepackage[version=4]{mhchem}

\usepackage{mathptmx}
\usepackage{siunitx}
\DeclareSIUnit{\atom}{at}

\begin{document}

\title{\fontsize{20}{26}\selectfont\bf Atomistic Mechanisms of Hard Carbon Formation\\ from Polyvinylidene Chloride}

\author[1]{Litong Wu}
\author[1]{Zitong Wu}
\author[1]{Zakariya El-Machachi}
\author[1]{Volker~L.~Deringer\thanks{volker.deringer@chem.ox.ac.uk}}

\affil[1]{Inorganic Chemistry Laboratory, Department of Chemistry, University of Oxford, Oxford, UK}
\date{}

\maketitle

\setstretch{1.5}

{\bf
Hard carbons are a class of disordered materials with widespread application in energy storage.
Despite decades of research, their atomistic formation mechanisms have remained elusive, due to the difficulty of both {\em in situ} experimental characterization and first-principles simulations.
Here, we describe the formation mechanism of hard carbon from the thermal decomposition of polyvinylidene chloride~(PVDC), using first-principles-quality simulations with a bespoke machine-learned interatomic potential model. 
Our results indicate a two-stage process, consisting of (i) radical-mediated dehydrochlorination, which generates reactive unsaturated carbon sites, and (ii) progressive carbon--carbon cross-linking followed by thermally activated rearrangement into an extended sp$^{2}$-bonded network. 
We provide an atomistic account of non-hexagonal ring motifs emerging during pyrolysis, supporting the empirically-derived theory that these motifs induce the intrinsic curvature that frustrates graphitic ordering in hard carbons.
}

\clearpage

\section*{Introduction}

In a multi-component device as intricate as a battery, achieving electrochemical compatibility across all components is crucial. 
The discovery of viable anode hosts has been a key challenge in this regard, as borne out by the historical development of both lithium-ion batteries~(LIBs) and sodium-ion batteries~(SIBs).~\cite{Wu-24-03, Winter-18-12a, Irisarri-15} 
The first commercial LIB, introduced in 1991, employed petroleum coke-derived soft carbon;~\cite{Megahed-94-08} subsequent work explored hard carbon synthesized from polyfurfuryl alcohol resin to achieve higher reversible capacity.~\cite{Sekai-93-03} 
Graphite later became the dominant LIB anode upon the recognition that its lithiation could be stabilized in ethylene-carbonate-based electrolytes through formation of a solid--electrolyte interphase.~\cite{Takehara-93-06, Yamaura-93-03, Xu-04-10} 
However, the situation is different in SIBs:~graphite does not form thermodynamically favorable Na intercalation compounds under conditions typical for batteries, whereas hard carbons can store Na ions reversibly within their disordered structure.~\cite{Ge-88-09, Stevens-01-06, Moriwake-17-07}
Since the first report of reversible Na-ion storage in glucose-derived hard carbon in 2000,~\cite{Stevens-00-04a} hard carbons have become the leading anode material for contemporary SIB technologies.~\cite{Guo-25-06, Wu-24-03} 

The recognition of hard carbons as a distinct class of disordered carbonaceous materials, structurally different from both crystalline graphite and fully amorphous carbon, can be traced back to early X-ray diffraction~(XRD) studies in the 1940s.~\cite{Biscoe-42-06, Gibson-46-01} 
The subsequent systematic distinction between hard and soft carbons (then termed as non-graphitizing and graphitizing carbons, respectively) was established through the foundational studies of Franklin.~\cite{Franklin-50-01, Franklin-51-05, Franklin-51-10}
Hard carbons are disordered materials that cannot be transformed into crystalline graphite even at temperatures as high as \SI{3000}{\celsius}, whereas soft carbons are disordered carbons that can.~\cite{Harris-25-10} 
These materials are typically synthesized by heating polymer precursors to approximately \SI{1000}{\celsius} in the absence of oxygen~(``pyrolysis''),~\cite{Devi-21-11} with the choice of precursor playing a central role in determining the nature of the resulting carbon.~\cite{Gibson-46-01, Uskokovic-21-10a}
A classic example is a pair of chlorinated polymers, polyvinylidene chloride~(PVDC) and polyvinyl chloride~(PVC), whose pyrolysis yields hard and soft carbons, respectively.~\cite{Franklin-51-10, Putman-18-08} 
As a first step towards understanding this contrast, we will here use PVDC as a model system to investigate how precursor structure directs hard-carbon formation.~\cite{Dacey-63-01, Ban-75-08, Pasek-96-01, Harris-97-09b, Kim-03-07}

Over the decades, experiments have provided key insights into the structure and formation of PVDC-derived hard carbon. 
Thermogravimetric analysis and mass spectrometry revealed a two-stage mass loss during PVDC pyrolysis: HCl evolution at around \SI{250}{\celsius}, and a minor release of hydrocarbon species near \SI{500}{\celsius}.~\cite{Endo-01-09, Kim-03-07} 
Infrared spectroscopy indicated that early carbonization involved the loss of \ce{C-Cl} functionalities, followed by the development of conjugated \ce{C=C} features.~\cite{Dacey-63-01} 
Electron spin resonance measurements showed paramagnetic signals during initial dehydrochlorination; this signal then diminished and semiconducting behavior emerged due to electron delocalization over the conjugated system that developed.~\cite{Winslow-55-09a, Hay-70} 
XRD studies confirmed that, although structural ordering increased with pyrolysis temperature, the crystallite dimensions along the $a$- and $c$-axes remained limited ($L_{a}$,~$L_{c}$\,$<$\,\SI{5}{nm}) even at~\SI{3000}{\celsius}. 
The persistence of a broad (002) reflection indicated a large and stable inter-layer spacing, in contrast to the layer contraction expected to be observed during graphitization.~\cite{Franklin-51-10, Ban-75-08, Kim-03-07, Putman-18-08}

The challenge, however, lies in translating these experimental signatures into a coherent atomistic understanding. 
Franklin attributed hard carbons' resistance to graphitization to randomly oriented graphitic domains, interconnected by covalent cross-links that impede rearrangement into a parallel structure, although the nature of these cross-links remains unclear.~\cite{Franklin-51-10} 
In the 1980s, Oberlin described the structural evolution in terms of forming and aligning basic structural units into locally oriented domains, with precursor-dependent viscosity governing domain growth and graphitizability.~\cite{Oberlin-84-01, Bonijoly-82-02} 
Following the discovery of fullerenes, Harris and Tsang proposed a model in which non-hexagonal rings introduce curvature into defective graphene-like fragments.~\cite{Harris-97-09b, Harris-97-09, Harris-00-06} 
This view was later supported by aberration-corrected transmission electron microscopy studies by Guo~{\em et~al.}\;(ref.~\citenum{Guo-12-08}) and Allen~{\em et~al.}\;(ref.~\citenum{Allen-22-02}).
Recently, Fogg~{\em et~al.}\;showed that PVDC-derived hard carbon can be transformed into highly crystalline graphite by repeated rapid heating pulses, proposing that selective removal of non-hexagonal defects enables graphitization.~\cite{Fogg-20-07}

While these models provide a structural framework for hard carbons, they do not explain how the proposed topological motifs arise from the reactive chemistry of a polymer precursor. 
Experimentally, chemical reactions are probed mainly through indirect and ensemble-averaged signatures. 
Computationally, resolving the reactive sequence requires an atomistic description of bond‐breaking and bond‐forming events across length and time scales beyond the reach of conventional first-principles methods.
Reactive molecular dynamics~(MD) simulations based on empirical force fields, particularly ReaxFF,~\cite{vanDuin-01-10} have provided valuable insight into the pyrolysis-driven carbonization of various polymer systems.~\cite{Purse-22-04, Kemppainen-23-10, Nahian-23-10a, Gallegos-25-03}
However, the accuracy and transferability of such empirical force fields remain constrained by their parameterization, particularly for chemically complex high-temperature pathways involving halogenated precursors. 
Machine-learned interatomic potentials~(MLIPs) offer a practical route forward by combining near first-principles fidelity with computational efficiency,~\cite{Behler-17-10, Deringer-19-11, Unke-21-08} both of which are important for describing the high-temperature structural evolution in chemically complex carbonaceous materials.
Recent polymer-focused datasets and foundation-model efforts further highlight the growing role of machine-learning approaches in atomistic polymer modeling.~\cite{Simm-25-10, Levine-25-12}

In the present work, we provide an atomistic account of the chemical and structural transformations underlying PVDC pyrolysis. 
We first characterize the radical-mediated dehydrochlorination cascade and examine how its regioselectivity evolves as the carbon network matures. 
We then identify a stepwise carbon cross-linking mechanism driven by the emergence of unsaturated carbon sites. 
Finally, we show how the initially defect-rich carbon cluster rearranges into a curved sp$^2$ network, in which the persistence of odd-membered rings frustrates planar graphitic ordering. 
These insights are enabled by a bespoke graph-based MLIP model together with a specialized MD protocol combining staged heating, volatile removal, and density-controlled cell rescaling. 
Our simulation connects precursor-level reaction chemistry to the emergence of topological disorder, providing an atomistic perspective on the non-graphitizing character of PVDC-derived hard carbon.

\clearpage

\section*{Results and Discussion}

\subsection*{Machine-Learning-Driven Simulation of PVDC Pyrolysis}

\begin{figure*}[t]
    \includegraphics[width=\linewidth]{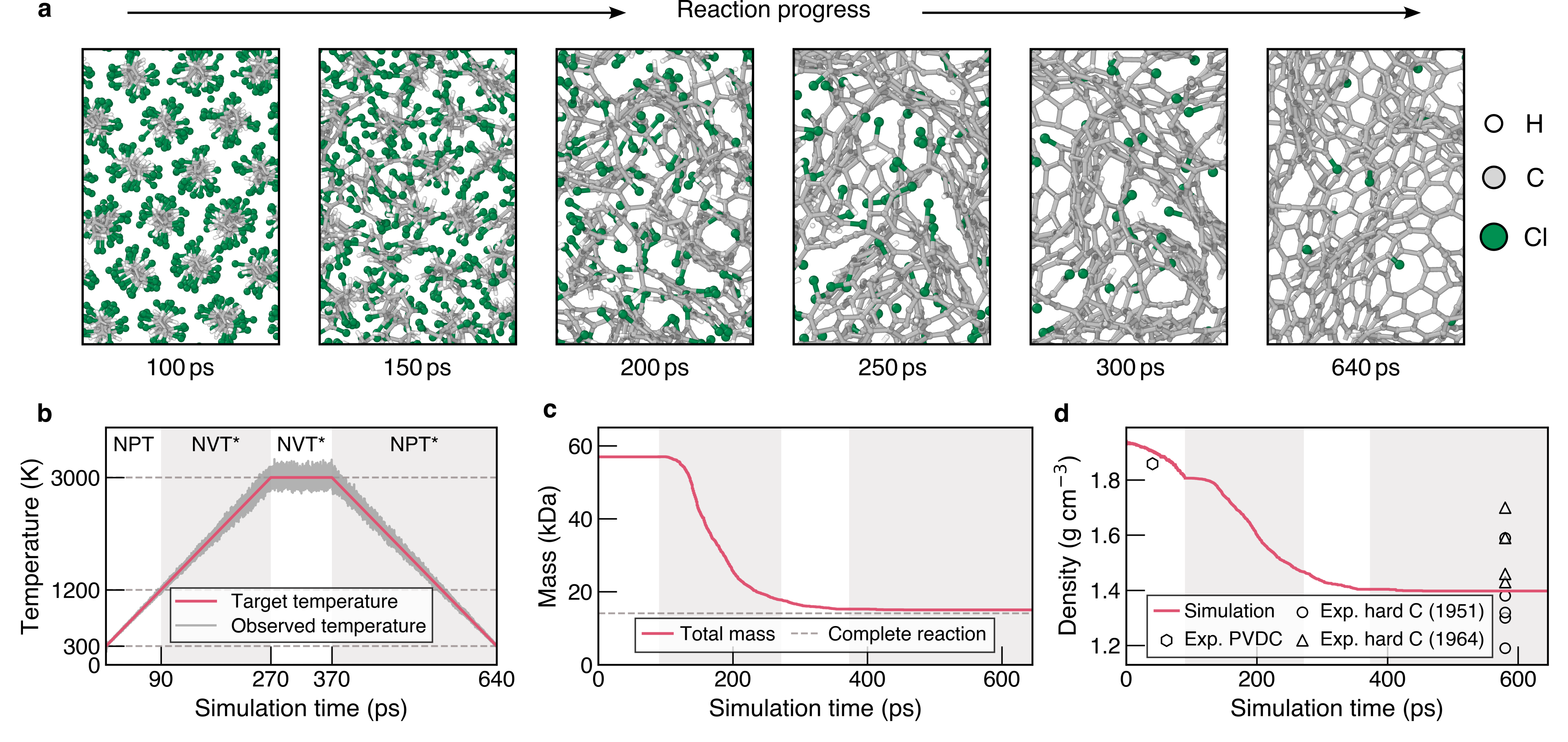}
    \caption{\textbf{Machine-learning-driven simulation of PVDC pyrolysis.} 
    (\textbf{a})~Representative snapshots from a \SI{640}{\pico\second} MD trajectory using a bespoke MLIP model. The initial snapshot shows PVDC viewed along the polymer-chain direction. The reaction proceeds via elimination of chlorine and hydrogen atoms as HCl molecules, followed by cross-linking of the resulting unsaturated carbon chains into a curved sp$^{2}$-rich network. 
    (\textbf{b})~Schematic overview of the MD protocol. The system was heated from \SI{300}{\kelvin} to \SI{3000}{\kelvin} at a rate of \SI{e13}{\kelvin\per\second}, annealed at \SI{3000}{\kelvin} for \SI{100}{\pico\second}, and quenched at the same rate. Ensembles used for each stage are indicated at the top; asterisks denote fixed-interval HCl removal combined with cell rescaling. The instantaneous temperature during the simulation is tracked in gray. 
    (\textbf{c})~Mass change associated with simulated HCl removal. The gray dashed line marks the mass expected after complete HCl elimination. 
    (\textbf{d})~Evolution of the system density. Experimental densities of commercial PVDC~(hexagon~\cite{Takahagi-88-10}) and PVDC-derived hard carbons~(circles~\cite{Franklin-51-10} and triangles~\cite{Kipling-64-04}) are shown for comparison. Background shading is used to highlight the different simulation stages defined in panel~(\textbf{b}), and is used in the same way in Figure~\ref{fig:2}a and \ref{fig:3}a.}
    \label{fig:1}
\end{figure*}

We performed MD simulations of PVDC pyrolysis with a bespoke MLIP fitted using the MACE architecture, \cite{Batatia-22-12} starting from a supercell constructed from the single-crystal XRD-refined structure.~\cite{Takahagi-88-10} 
Details of the dataset and model-training procedure are described in the Methods section, while numerical validation and robustness tests are presented in Section~S1.2 of the Supporting Information.
Representative snapshots of the system throughout the pyrolysis process, viewed along the polymer-chain direction, are shown in Figure~\ref{fig:1}a. 
These snapshots provide a microscopic view of the transformation: as the temperature increases, dehydrochlorination generates unsaturated carbon sites that subsequently cross-link and progressively coalesce into a disordered hard-carbon network.

Figure~\ref{fig:1}b illustrates the MD protocol, with the instantaneous simulation temperature superimposed on the target temperature profile.
The system was heated from 300 to \SI{3000}{\kelvin} at a rate of \SI{e13}{\kelvin\per\second}, with a switch from the NPT to the NVT$^\ast$ ensemble at \SI{1200}{\kelvin} to avoid simulation-cell instability caused by the rapid evolution of gaseous molecules at higher temperature. 
(We use asterisks to indicate that HCl molecules were periodically removed and the simulation cell isotropically rescaled to maintain a physically realistic density; see Methods.)
The system was then annealed at \SI{3000}{\kelvin} for \SI{100}{\pico\second} and quenched back to \SI{300}{\kelvin} in the NPT$^\ast$ ensemble, again at a rate of \SI{e13}{\kelvin\per\second}. 
The four stages of the MD protocol are indicated by alternating background shading in our plots.

Given that PVDC consists of alternating hydrogen and chlorine substituents along the polymer backbone, its primary thermal decomposition pathway is dehydrochlorination:
\begin{center}
    \includegraphics[width=0.4\linewidth]{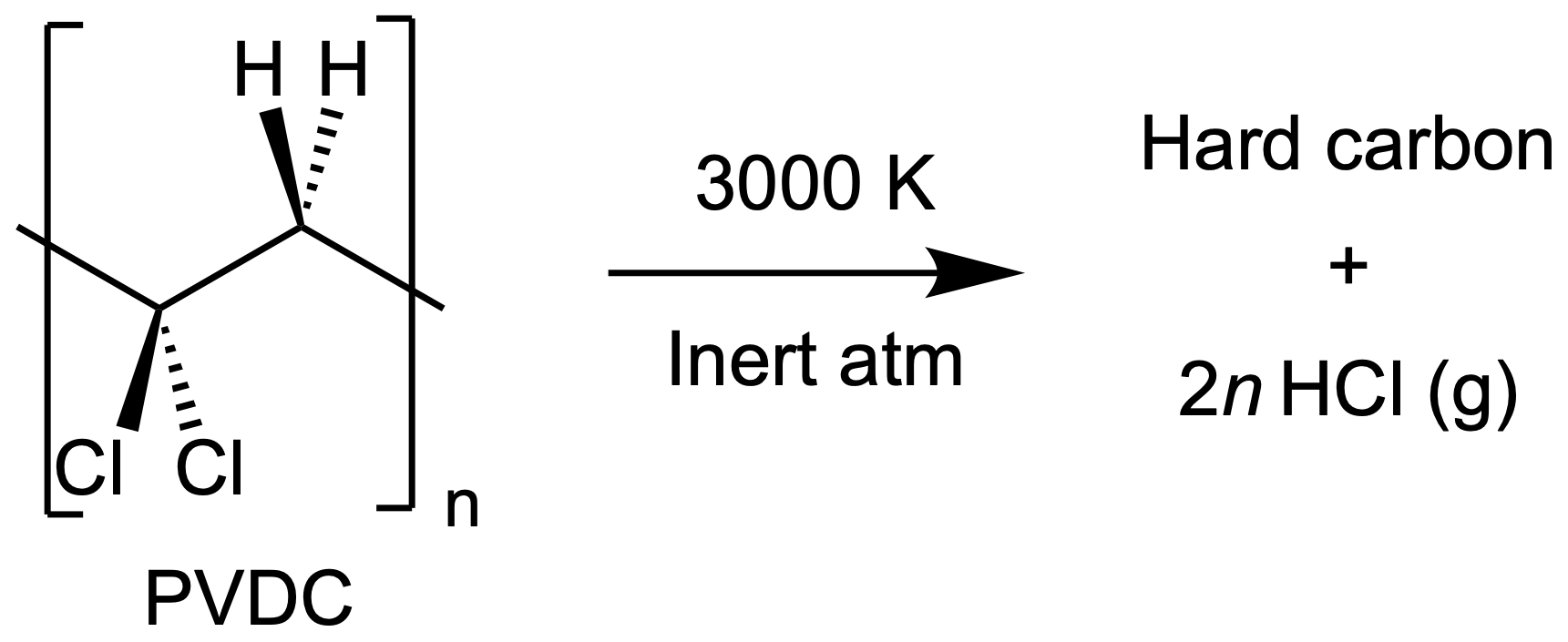}
\end{center}
\vspace{-0.8em}
This behavior is well captured by our simulation: HCl accounts for 83.1\% of all molecular products detected and is the only product removed (Supporting Information,~Section~S2.1). 
Although transient \ce{H2} and \ce{Cl2} molecules are occasionally formed, their low abundance is consistent with experimental studies identifying HCl as the principal volatile product of PVDC degradation.~\cite{Davies-71-01, Montaudo-91-01, Endo-01-09, Kim-03-07}

The resulting cumulative mass evolution due to the removal of HCl molecules formed is plotted in Figure~\ref{fig:1}c. 
The first HCl removal occurs after \SI{100}{\pico\second}, when the simulation temperature reaches approximately \SI{1300}{\kelvin}. 
Since all HCl molecules produced within each 1-ps interval are removed at the end of that interval during the NVT$^\ast$ stages, the resulting mass profile provides a measure of the reaction kinetics. 
Dehydrochlorination appears to be consistent with pseudo-first-order kinetics under the imposed heating protocol, where the reaction rate decreases as the available H- and Cl-containing sites are consumed (Supporting Information,~Section~S2.2). 
The last HCl removal occurred during quenching, when the system reached \SI{2200}{\kelvin}, by which point 97.9\,\% of the total HCl had been eliminated. 

To assess the physical validity of the MD protocol, the density evolution is presented in Figure~\ref{fig:1}d. 
The initial density of the simulation cell is slightly higher than that typically reported for bulk PVDC,~\cite{Takahagi-88-10} because the starting structure is derived from a defect-free single-crystal XRD model. 
During the NPT heating phase, the density decreased slightly due to thermal expansion. 
Throughout the NVT$^\ast$ stages, the density is governed by the rescaling parameters and remains constant over the first \SI{10}{\pico\second}, as no HCl has yet been eliminated. 
A moderate decrease in density was observed at \SI{100}{\pico\second}, concomitant with the onset of dehydrochlorination, and proceeded at a progressively slower rate that mirrors the declining reaction rate.
At the onset of the quenching stage, despite switching to an NPT$^\ast$ ensemble at \SI{3000}{\kelvin}, the density remained approximately constant at \SI{1.4}{\gram\per\cubic\cm}. 
This behavior indicates that the carbon network has already developed substantial mechanical rigidity, consistent with earlier experimental evidence that dehydrochlorination-generated polyene chains undergo early inter-chain cross-linking which effectively ``freezes'' the structure.~\cite{Davies-71-01} 
As the density evolution under this simulation protocol is sensitive to the chosen rescaling parameter, the parameter was calibrated such that the final density of the simulated hard carbon falls within the reported range for PVDC-derived hard carbons.~\cite{Franklin-51-10, Kipling-64-04}

\subsection*{Dehydrochlorination Mechanism}

\begin{figure*}[p]
    \includegraphics[width=\linewidth]{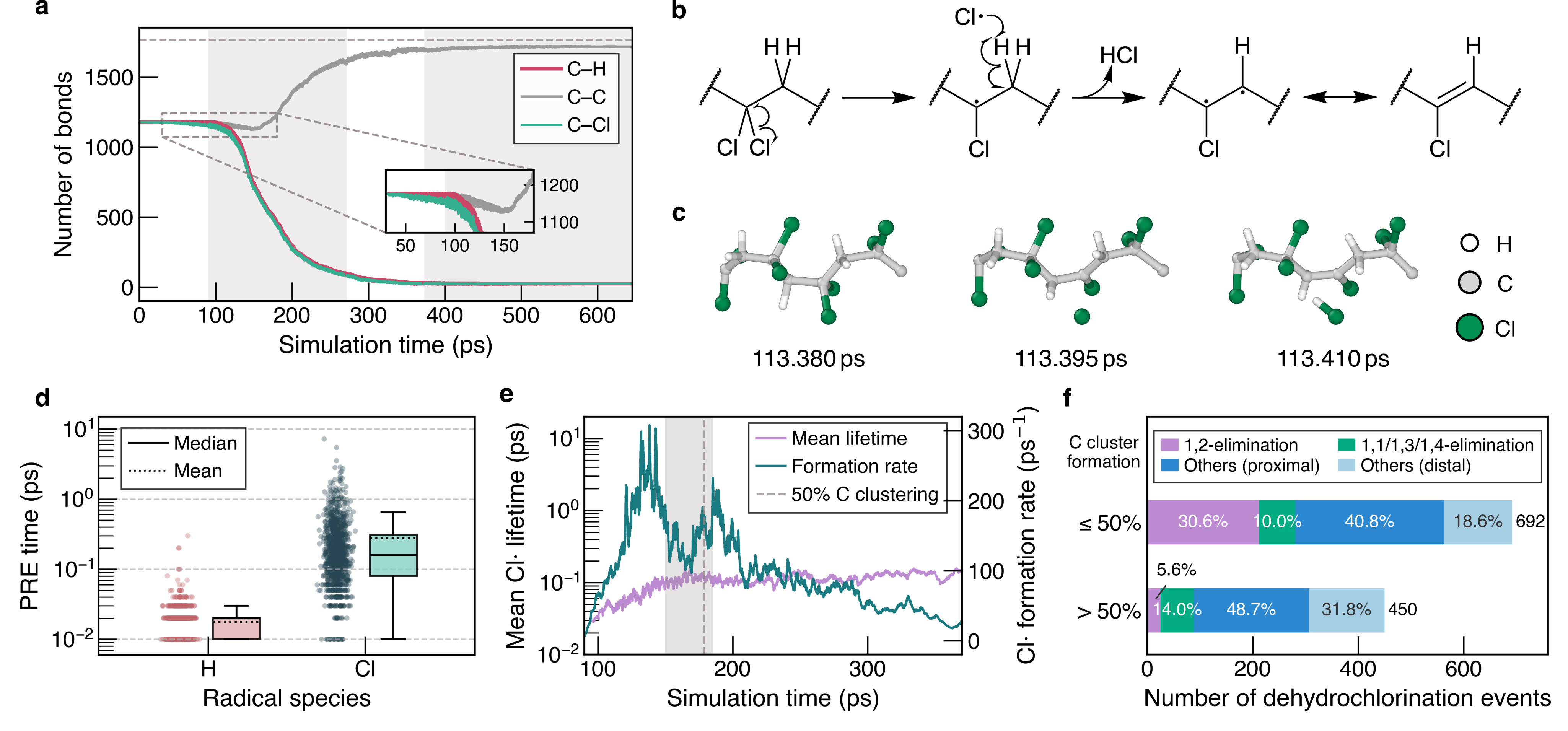}
    \caption{\textbf{Radical-mediated dehydrochlorination mechanism.} 
    (\textbf{a})~Evolution of \ce{C-H}~({\em pink}), \ce{C-C}~({\em grey}), and \ce{C-Cl}~({\em green}) bond counts during the simulation. 
    The inset shows the 30--180\,ps interval, highlighting that \ce{C-Cl} cleavage precedes \ce{C-H} cleavage, and that \ce{C-C} scission occurs prior to carbon-network growth. 
    (\textbf{b})~Schematic of the proposed radical dehydrochlorination mechanism, proceeding via homolytic \ce{C-Cl} bond cleavage, hydrogen abstraction by \ce{Cl.} radicals, and \ce{C-C} $\pi$-bond formation. 
    (\textbf{c})~Representative fragments from the simulation trajectory illustrating key steps of the proposed mechanism. 
    (\textbf{d})~Pre-reactive-encounter~(PRE) times of eliminated H~({\em pink}) and Cl~({\em green}) atoms. 
    For each HCl-elimination event, the time is defined as the interval between the final detachment of the atom from the carbon framework and its first approach within \SI{1.50}{\angstrom} of any complementary species.
    Each dot represents an individual atom, with horizontal jitter added for visual clarity. 
    Box plots span the interquartile range (IQR), with solid and dotted lines denoting the median and mean, respectively; whiskers extend to \(1.5\times\) the IQR.
    (\textbf{e})~Evolution of the average lifetime~({\em purple}) and formation rate~({\em green}) of \ce{Cl.} radicals defined on an event basis during the NVT$^\ast$ stages~(\SIrange{90}{370}{\pico\second}). 
    The point at which 50\% of the C atoms have joined the largest connected carbon cluster is indicated by a dashed line, and the clustering period is shaded.
    (\textbf{f})~Regioselectivity of HCl elimination before and after 50\% carbon-cluster formation. Events are classified according to the relative positions of the parent carbon atoms as 1,2-elimination~({\em purple}), 1,1/1,3/1,4-elimination~({\em green}), and other eliminations as described in the main text.}
    \label{fig:2}
\end{figure*}

Having characterized the macroscopic evolution of the system during pyrolysis, we next analyze the bond-level changes underlying dehydrochlorination. 
An examination of the bond-count evolution in Figure~\ref{fig:2} provides insight into the mechanism. 
The overall decrease in \ce{C-Cl} and \ce{C-H} bond counts is consistent with the observed mass loss due to HCl removal, while the increase in \ce{C-C} bonds at around \SI{150}{\pico\second} suggests the onset of carbon cross-linking. 
However, the expanded view in the inset of Figure~\ref{fig:2}a reveals that the cleavage of \ce{C-Cl} bonds precedes that of \ce{C-H} bonds; the time at which 1\% of all \ce{C-Cl} bonds had remained dissociated for longer than \SI{0.2}{\pico\second} preceded that for \ce{C-H} bonds by \SI{42.4}{\pico\second}.
In addition, the \ce{C-C} bond count undergoes a shallow decline by 4.4\% at \SI{146.4}{\pico\second} before increasing again, implying thermally induced or radical-driven \ce{C-C} scission prior to net carbon-network growth.

These observations are consistent with a radical-mediated dehydrochlorination mechanism, as illustrated schematically in Figure~\ref{fig:2}b. 
The process is initiated by thermally induced \ce{C-Cl} bond homolysis, which generates reactive \ce{Cl.} radicals capable of abstracting hydrogen. 
When abstraction occurs from an adjacent carbon atom, HCl elimination is likely coupled to local \ce{C=C} formation, as illustrated in Figure~\ref{fig:2}b.
However, if the \ce{Cl.} radical migrates further before hydrogen abstraction, the resulting under-coordinated carbon sites are spatially separated from the original cleavage site and can participate in competing processes, including further chlorine detachment, radical migration, or termination through inter-chain \ce{C-C} bond formation. 
Thus, Figure~\ref{fig:2}b should be regarded as a simplified representation of the local HCl-forming pathway, rather than a complete description of all dehydrochlorination events observed during pyrolysis.
Representative structural fragments extracted from the simulation trajectory, capturing key instances of these bond-level events, are shown in Figure~\ref{fig:2}c.
In the following trajectory analysis, the term ``radical'' is used operationally to denote a detached species identified based on bond connectivity, rather than an electronically characterized radical state: 
although the DFT reference computations used for MLIP training were spin-polarized, the connectivity-defined radical species identified here do not necessarily imply persistent localized unpaired-spin states.

Figure~\ref{fig:2}d shows the pre-reactive-encounter~(PRE) times for isolated hydrogen and chlorine species. 
For each eliminated HCl molecule, we define this quantity as the time interval between the last instance at which either the hydrogen or chlorine atom is bonded to a carbon atom, and the subsequent moment this atom first enters the reactive domain of a complementary species. 
Defining the time origin with respect to the last \ce{C-X} (X\,$=$\,H~or~Cl) bond is necessary because radicals formed during the simulation frequently recombine to the carbon network before their final removal. 
The PRE time is also intended to capture the duration for which a detached species remains isolated before its first reactive encounter; subsequent partner-exchange events prior to final HCl elimination are not included. 
Moreover, a more permissive \ce{H-Cl} cutoff of \SI{1.50}{\angstrom} is used to define the reactive domain, rather than the \SI{1.30}{\angstrom} cutoff employed as the criterion for molecular removal. 
This separation of criteria accounts for high-temperature bond-length fluctuations, where \ce{H-Cl} pairs may already be associated at thermally elongated separations before contracting below the removal cutoff. 

The average PRE time for \ce{Cl.} radicals is significantly longer than that for hydrogen, implying that the former persist as isolated species while diffusing to reactive hydrogen sites. 
In contrast, hydrogen atoms remained isolated for less than \SI{0.02}{\pico\second} on average, a timescale comparable to the period of a single \ce{C-H} bond vibration (\SI{0.01}{\pico\second} based on a wavenumber of \SI{3000}{\per\cm}).~\cite{Buck-01-11a} 
This likely suggests that isolated hydrogen atoms are not stabilized as independent radicals but exist only as transient states during abstraction. 
These disparate timescales support the proposed mechanism, in which the \ce{Cl.} radicals act as the primary mobile species in the dehydrochlorination process.

To further characterize the temporal behavior of \ce{Cl.} radicals, we analyzed their lifetime using an event-based algorithm. 
In this approach, each homolytic bond dissociation is treated as an independent event regardless of the atom's history: if a \ce{Cl.} radical re-binds to the carbon network and subsequently detaches again, the newly formed radical is recorded as a separate data point. 
The resulting ensemble-averaged lifetimes are presented as a function of their generation time in purple in Figure~\ref{fig:2}e. 
The average lifetime of \ce{Cl.} radicals increases during the initial stage and stabilizes at approximately \SI{0.10}{\pico\second} around the \SI{170}{\pico\second} mark. 
This value is lower than the mean PRE time shown in Figure~\ref{fig:2}d, as it also accounts for radical events that do not lead to HCl elimination.

The formation rate of \ce{Cl.} radicals, as a function of their generation time, is shown by the green line in Figure~\ref{fig:2}e. 
This quantity increases initially, reaching a maximum at \SI{138.33}{\pico\second}. 
Although this rate subsequently declines in tandem with the consumption of chlorine content, a secondary peak emerges at \SI{185.22}{\pico\second}, following a local minimum at \SI{169.51}{\pico\second}. 
This bimodal profile suggests that dehydrochlorination likely proceeds through two kinetically distinct stages under the imposed heating protocol. 
The later stage may involve elimination from already-unsaturated local environments, where higher activation barriers are accessed only at elevated temperature.

In general, the early stage of dehydrochlorination is characterized by rapid generation of \ce{Cl.} radicals with relatively short lifetimes, whereas at later stages the radical population decreases while the average lifetime increases. 
This apparent inverse relationship can be understood from the evolving regioselectivity of HCl elimination, as shown in Figure~\ref{fig:2}f. 
The relative positions of the parent carbon atoms associated with each eliminated HCl pair are computed and partitioned into two stages based on the development of the carbon framework -- before and after the point at which 50\% of the carbon atoms have joined the largest connected carbon cluster (indicated by a gray dashed line in Figure~\ref{fig:2}e). 
We categorize elimination events as 1,2-elimination, 1,1/1,3/1,4-elimination, or others. 
Although chlorine and hydrogen atoms are initially bonded to alternating carbon atoms in the precursor polymer, 1,1- and 1,3-eliminations occur due to frequent radical recombination and cross-linking, which alter the local atomic environment. 
These non-standard eliminations support the radical-mediated nature of the reaction. 
The ``others'' category, comprising long-range intra-chain and any inter-chain eliminations, is further subdivided based on the direct distance between parent carbon atoms. 
A threshold of \SI{6.7}{\angstrom}, corresponding to the mean distance across all such events, is used to distinguish between proximal and distal eliminations.

Overall, the majority of dehydrochlorination events (60.5\%) occurs prior to 50\% carbon-cluster development. 
During this early stage, 1,2-elimination is one of the dominant pathways, favored by the intrinsic proximity of adjacent reactive sites, which allows for efficient abstraction with minimal radical diffusion. 
Concurrently, the proportion of proximal inter-chain elimination is significant, which can be attributed to the crystalline packing of the precursor; 
specifically, the minimum inter-chain Cl$\cdots$H distance of \SI{3.22}{\angstrom} is substantially shorter than the maximum 1,4-intra-chain distance of \SI{6.20}{\angstrom}, making inter-chain abstraction spatially favorable compared to long-range intra-chain events.
This behavior is consistent with the classical picture of competing zip-like intra-chain propagation and inter-chain radical transfer.~\cite{Davies-71-01}

As carbon cross-linking progresses, the proportion of 1,2-elimination diminishes in favor of distal and inter-chain pathways. 
Taken together with the increasing \ce{Cl.} lifetime in Figure~\ref{fig:2}e and the higher, more heterogeneous apparent chlorine mobility at later times (Supporting Information, Section~S2.3), this shift suggests that HCl formation becomes increasingly encounter-limited as nearby H-containing sites are depleted and the carbon framework rigidifies.

\subsection*{Carbon-Network Formation}

\begin{figure*}[t]
    \includegraphics[width=\linewidth]{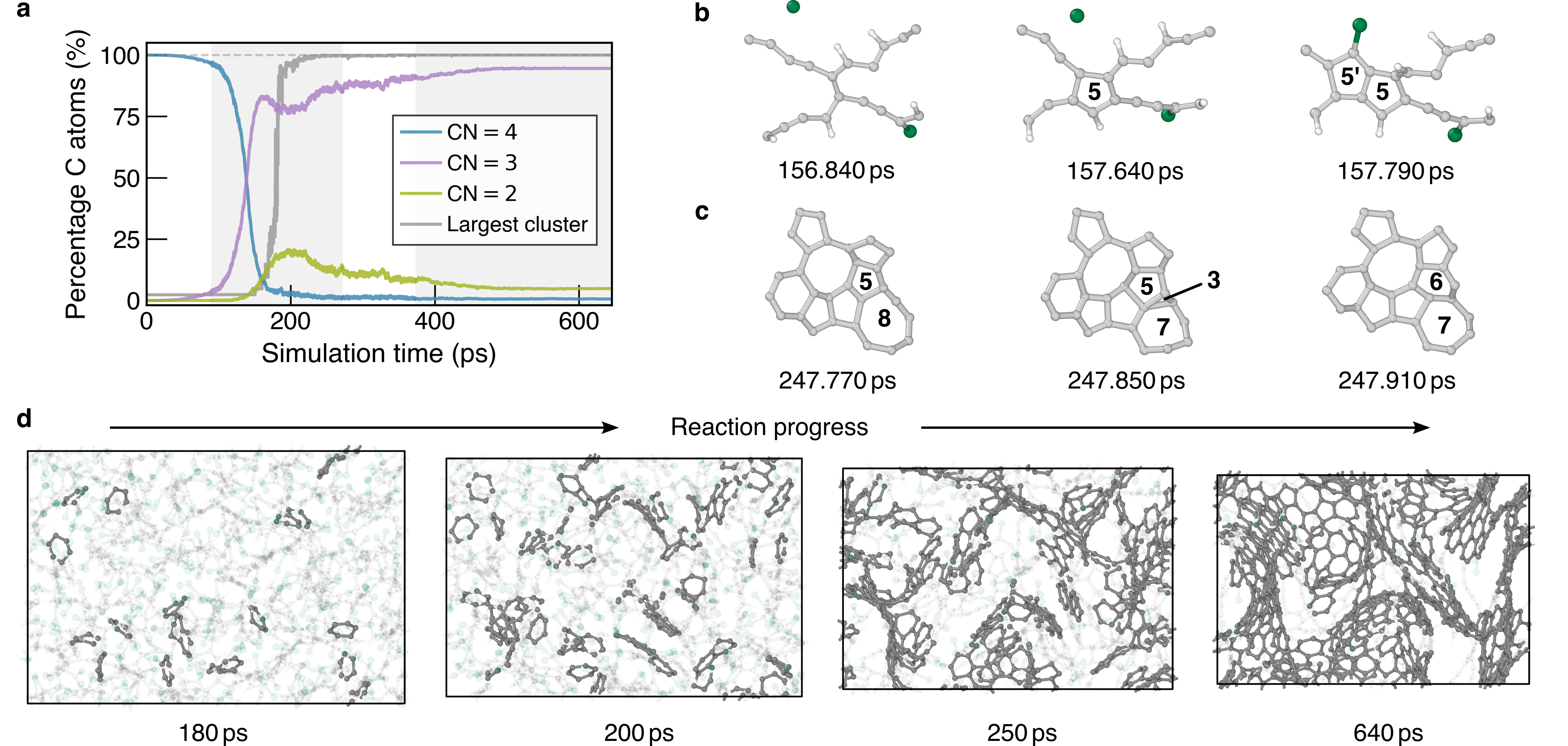}
    \caption{\textbf{Carbon network formation and rearrangement.} 
    (\textbf{a})~Evolution of the relative abundance of carbon atoms categorized by coordination number~(CN):~four~({\em blue}), three~({\em purple}), and two~({\em green}). The gray line tracks the fraction of carbon atoms belonging to the largest cluster, defined as the largest set of carbon atoms linked through continuous \ce{C-C} bonds.
    (\textbf{b})~Representative fragments illustrating the stepwise cross-linking of unsaturated carbon chains and the resulting ring formation.
    (\textbf{c})~Representative fragments illustrating the rearrangement of five- and eight-membered rings into six- and seven-membered rings, via a three-membered ring intermediate. 
    (\textbf{d})~Structural snapshots highlighting six-membered rings to track their spatial development throughout the simulation. Frames are drawn to scale with the simulation cell.}
    \label{fig:3}
\end{figure*}

Having examined the mechanisms of dehydrochlorination, we now turn to the structural evolution of the carbon network. 
Figure~\ref{fig:3}a depicts the temporal evolution of carbon atoms categorized by their coordination numbers~(CNs). 
Notably, an initial decrease in the population of fourfold-coordinated carbon is observed before the first HCl-removal event at \SI{100}{\pico\second}. 
We attribute this structural change to the homolytic cleavage of \ce{C-Cl} bonds, the accumulation of which sets the stage for subsequent dehydrochlorination. 
Beyond \SI{100}{\pico\second}, the rapid conversion of fourfold- to threefold-coordinated carbon atoms reflects the accelerating rate of dehydrochlorination.

Given that each carbon atom in pristine PVDC is bonded to two substituents, continued dehydrochlorination would be expected to drive a further transition from threefold- to twofold-coordinated states. 
Indeed, the $\text{CN}=3$ population initially rises to 83.6\% at \SI{158.77}{ps}, then declines to 75.9\% at \SI{196.32}{ps}, where the $\text{CN}=2$ population reaches its maximum. 
This behavior coincides with the second maximum in the \ce{Cl.} formation rate in Figure~\ref{fig:2}e, which occurs at \SI{185.22}{\pico\second}, during the same transition from threefold- to twofold-coordinated carbon.
However, most of these $\text{CN}=2$ sites are quickly consumed, as their appearance is followed by a sharp rise in the size of the largest carbon cluster, defined as the maximal set of carbon atoms connected via continuous \ce{C-C} bonds. 
This suggests simultaneous \ce{C-C} network development, where these high-energy twofold-coordinated intermediates are rapidly incorporated into the growing network, restoring a more stable sp$^2$-rich state.
The MACE-predicted per-atom energy contributions for carbon atoms with different local CN are given in Section~S2.6 of the Supporting Information, where twofold-coordinated carbon atoms are shown to be associated with the highest local energies among the coordination environments analyzed.

Selected structural fragments in Figure~\ref{fig:3}b illustrate the cross-linking of carbon chains through a stepwise mechanism mediated by twofold-coordinated carbon intermediates. 
These under-coordinated sites are highly reactive and enable successive \ce{C-C} coupling events between neighboring unsaturated chains.
The time separation between successive \ce{C-C} bond-formation events is on the order of picoseconds, much longer than typical \ce{C-C} vibrational periods (\SIrange{20}{50}{\femto\second} for stretching modes and up to \SI{100}{\femto\second} for bending modes),~\cite{Buck-01-11a} indicating that ring closure proceeds stepwise rather than in a concerted fashion.
We also note that the \ce{Cl.} radical visible in the upper left of the first two frames in Figure~\ref{fig:3}b recombines with the carbon network in the final frame, exemplifying a radical-addition event that occurs frequently throughout the simulation.

Although carbon-network aggregation is rapid and reaches a plateau by \SI{185.17}{ps}, the structural snapshot at around this stage (first frame in Figure~\ref{fig:3}d) contains only a sparse population of six-membered rings. 
This observation suggests that, although the carbon scaffold assembles rapidly, aromatic ordering continues through subsequent internal rearrangement. 
The evolution of five- to seven-membered ring populations (Figure~S6a) shows that five- and six-membered rings form rapidly at early stages of the simulation, whereas seven-membered rings emerge more slowly. 
We also note a slight decrease in the seven-membered-ring population and a continued increase in six-membered rings after five-membered-ring formation has saturated, suggesting continued annealing of larger non-hexagonal motifs into more stable six-membered rings during early quenching. 
One such pathway is shown in Figure~\ref{fig:3}c, where a fused five- and eight-membered ring pair undergoes stepwise conversion into fused six- and seven-membered motifs, via a three-membered-ring intermediate. 

However, as shown in Figure~S6a, five- and seven-membered rings remain substantial in the final structure even after extensive quenching. 
Figure~\ref{fig:3}d, which highlights six-membered rings in the trajectory snapshots, shows their progressive growth towards the end of pyrolysis within a curved network. 
This curvature is linked to the presence of odd-membered rings, as shown in Figure~S6c, where threefold-coordinated carbon atoms participating in fewer six-membered rings exhibit larger out-of-plane distances. 
In addition, these threefold-coordinated carbon atoms are also associated with higher atomic energies (Figure~S8b) -- although the energetic penalty appears to be modest enough to allow them to persist in the final structure. 
Taken together, these results suggest that five- and seven-membered rings are not merely transient defects, but are incorporated into the evolving carbon framework in a way that preserves local curvature and frustrates complete graphitic ordering.

\section*{Conclusions}

We have developed a bespoke MLIP model and a density-calibrated simulation protocol to investigate the atomistic mechanisms of hard carbon formation from PVDC decomposition. 
Our results show that the carbonization stage of pyrolysis proceeds via thermally induced, radical-mediated dehydrochlorination, initiated by homolytic \ce{C-Cl} bond cleavage and the formation of reactive \ce{Cl.} species. 
This mechanism is similar to that proposed by Davies~{\em et~al.}\;on the basis of kinetic measurements,~\cite{Davies-71-01} and is compatible with recent quantum-mechanical nanoreactor simulations of PVC pyrolysis that likewise identified chlorine-radical-initiated hydrogen abstraction.~\cite{Lei-25-05}
The regioselectivity of subsequent hydrogen abstraction evolves with the progress of pyrolysis, with \ce{Cl.} radicals diffusing over longer distances to react as the network matures. 
In particular, we identify secondary elimination from already-unsaturated sites as a key process in driving a stepwise carbon cross-linking process. 
This leads to the formation of a defect-rich carbon network, which is subsequently annealed through internal rearrangement. 
The persistence of five- and seven-membered rings supports the long-standing view that non-hexagonal defects introduce curvature into graphene-like fragments.~\cite{Harris-97-09b, Harris-97-09, Harris-00-06, Allen-22-02, Guo-25-06} 
Our work complements this structural picture by providing an atomistic pathway for the formation of these defects during PVDC pyrolysis.
 
Two directions for further study therefore emerge.
First, MLIP-driven simulations offer a compelling opportunity to systematically compare PVC and PVDC pyrolysis, which could provide an atomistic basis to revisit Franklin's foundational classification of hard and soft carbon materials.~\cite{Franklin-50-01, Franklin-51-05, Franklin-51-10} 
Although the MLIP model developed in this work contains the necessary chemical elements, a quantitative PVC study would require additional training  and validation on PVC-specific environments. 
Furthermore, the MD protocol would need to be adapted to accommodate the more complex volatile-product chemistry of PVC degradation, since the hydrogen and chlorine contents are no longer present in stoichiometric balance and HCl is therefore no longer expected to be the only dominant volatile product.
Second, the carbon structures generated here could serve as realistic starting points for simulating Li- and Na-ion insertions, extending beyond prior studies that relied on melt–quench-derived proxies.~\cite{Deringer-18-05, Huang-19-07, Front-26-01} 
In a wider context, our study illustrates the growing role of atomistic ML methods in understanding the chemical mechanisms of the synthesis of structurally complex functional materials, paving the way for predictive simulations that guide the design of amorphous materials at the atomic level.

\clearpage

\setstretch{1.2}

\section*{Methods}

{\bf Dataset.} The final MLIP model was trained on a dataset comprising 3,726 structures (426,662 atoms). 
Most PVDC-pyrolysis configurations were generated through active learning. 
A total of 1,172 structures were produced using the CASTEP$+$ML framework, following Ref.\!\citenum{Stenczel-23-07}. 
This approach accelerates {\em ab initio} MD simulations by fitting Gaussian Approximation Potential~(GAP)~\cite{Bartok-10-04} models on the fly, with uncertainty-based criteria determining whether forces are evaluated using DFT or ML predictions; this framework has been used previously in the construction of a dataset and MLIP model for graphene oxide.~\cite{El-Machachi-24-12} 
An additional 957 PVDC-based structures were sampled from iterative MD trajectories. 
These simulations were performed on 216- or 192-atom supercells derived from single-crystal XRD data~\cite{Takahagi-88-10} and encompassed a broad range of thermodynamic conditions: 
(i)~NVT simulations up to \SI{3000}{\kelvin}; 
(ii)~NPT simulations at \SI{1}{\bar} and up to \SI{1800}{\kelvin}; 
(iii)~NPT compression simulations spanning \SIrange{1000}{3000}{\kelvin} and \SIrange{10}{150}{\giga\pascal}; 
and (iv)~trial pyrolysis simulations. 
To improve coverage of molecular by-products, the dataset was augmented with 170 chlorinated hydrocarbon molecules from PubChem~\cite{Kim-25-01} and their distorted variants, as well as 358 SOAP~\cite{Bartok-13-05}-selected QM9~\cite{Williams-25-01} molecules featuring random Cl-for-H substitutions with scaled bond lengths. 
The description of bulk carbon phases was reinforced with 450 amorphous carbon,~\cite{Rowe-20-07a} 162 graphite,~\cite{Rowe-20-07a} and 457 \ce{C-H}~\cite{Ibragimova-25-02} configurations from previous studies.

{\bf DFT computations.} All structures in the dataset were relabeled for consistency. 
Single-point energies, forces, and stresses were computed using density functional theory~(DFT) as implemented in CASTEP~24.1,~\cite{Stewart-05-02} with the Perdew--Burke--Ernzerhof~(PBE) generalized-gradient approximation~\cite{Perdew-96-10} and Tkatchenko--Scheffler~(TS) semi-empirical dispersion corrections.~\cite{Tkatchenko-09-02} 
The plane-wave cutoff energy was set to \SI{700}{\electronvolt}, with a self-consistent field convergence threshold of \SI{e-7}{\electronvolt}. 
To accurately describe open-shell species generated during bond dissociation, spin-polarized computations were performed; 
an initial magnetic moment of $1\,\mu_{\mathrm{B}}$ was assigned to each atom and allowed to relax freely to facilitate convergence. 
The Brillouin zone was sampled using Monkhorst--Pack grids~\cite{Monkhorst-76-06} with a maximum $k$-point spacing of $2\pi\,\times\,$\SI{0.04}{\per\angstrom}. 
Core electrons were described using on-the-fly-generated pseudopotentials in CASTEP. 

{\bf MLIP model.} We trained a bespoke MACE potential model~\cite{Batatia-22-12} to describe the thermal decomposition of PVDC. 
The dataset was partitioned by assigning 10\% of each configuration type to the test set and a further 10\% of the remaining structures to the validation set. 
We employed a \texttt{ScaleShiftMACE} architecture comprising two message-passing layers with 128 equivariant channels and equivariant features up to $l=1$. 
Each interaction block used a correlation order of three and included angular channels up to \(l_{\max}=3\), with a cutoff radius of \(r_{\max}=\SI{6.0}{\angstrom}\).
Training was conducted using a standard weighted energy--forces loss function, with loss weights set to $\lambda_{E}\,=\,1$, $\lambda_{F}\,=\,100$ for the first 150 epochs, and adjusted to $\lambda_{E}\,=\,1000$, $\lambda_{F}\,=\,100$ for the final 100 epochs. 
The model was trained using the Adam optimizer~\cite{Kingma-14-12} with a batch size of 6, an initial learning rate of~0.01, a stochastic weight-averaging~(SWA) learning rate of~0.001, and weight decay of $5 \times 10^{-7}$. 
While the MACE model itself was trained only on energies and forces, it can reproduce the spin-ground-state potential-energy surface locally. 
On the test set, our final model achieved root-mean-square errors of \SI{6.9}{\milli\electronvolt\per\atom} (\SI{0.67}{\kilo\joule\per\mole}) for energies and \SI{269.6}{\milli\electronvolt\per\angstrom} for forces. 
All training was performed on a single RTX 6000 Ada GPU.

{\bf MD simulations.} MD simulations were performed using LAMMPS~\cite{Thompson-22-02} with the \texttt{MACE/kk} pair style in KOKKOS mode on a single GPU. 
The simulation protocol is summarized in Figure~\ref{fig:1}b and discussed in the Results section. 
To maintain numerical stability and resolve high-temperature atomic motion, the integration timestep was adjusted according to the simulation stage: 
\SI{0.5}{\femto\second} for the NPT heating, \SI{0.25}{\femto\second} for the two NVT$^\ast$ stages, and \SI{1.0}{\femto\second} for the NPT$^\ast$ quenching. 
A Nos{\'e}--Hoover thermostat and barostat~\cite{Nose-83-11} were employed, with temperature and pressure damping constants set to \SI{0.1}{\pico\second} and \SI{1.0}{\pico\second}, respectively. 
The initial simulation supercell comprised 42 parallel PVDC chains (3528 atoms), with dimensions of \SI{47.0}{\angstrom}$\,\times\,$\SI{34.0}{\angstrom}$\,\times\,$\SI{31.5}{\angstrom}. Periodic boundary conditions were applied in all three spatial directions.

Starting from the second stage, segmented MD simulations were performed using a custom Python workflow, which automated periodic HCl removal, cell rescaling, and the initiation of successive simulation segments. 
The NVT$^\ast$ stages and the final NPT$^\ast$ quench were partitioned into segments of \SI{1.0}{\ps} and \SI{10.0}{\ps}, respectively. 
During each segment, the simulation advanced according to the programmed temperature ramp, isothermal hold, or quench. 
Between successive segments, the simulation was paused to identify and remove HCl molecules within a bonding cutoff of \SI{1.30}{\angstrom}. 
To maintain a physically realistic density following the volatile loss, the simulation cell and the atomic coordinates were isotropically rescaled. 
The updated structure was written to a LAMMPS data file along with velocities and used to initialize the subsequent segment.

The volume contraction associated with the removal of each HCl molecule was estimated to be around \SI{29.7}{\cubic\angstrom} from the sum of the spherical van der Waals volumes of H ($r_{\text{vdW}} = $\SI{1.20}{\angstrom}) and Cl ($r_{\text{vdW}} = $\SI{1.75}{\angstrom}) atoms.~\cite{Bondi-64-03}
We recognize, however, that the volume reduction associated with HCl removal in a condensed pyrolyzing solid is not necessarily captured by the sum of isolated van der Waals volumes; 
this quantity was thus treated as an empirical parameter that ought to be calibrated against experimental density data.
A value of \SI{30.0}{\cubic\angstrom} per removed HCl molecule was adopted, as it yielded good agreement with reported densities for PVDC-derived hard carbons.
For comparison, a larger value of \SI{33.5}{\cubic\angstrom} was also tested, which resulted in a higher final density and an increased fraction of threefold-coordinated carbon atoms, while preserving the qualitative features of the pyrolysis process. 

A detailed analysis of the sensitivity of the simulated pyrolysis pathway to the density-rescaling parameter and the maximum annealing temperature can be found in Section~S1.3 of the Supporting Information.

{\bf Structural analysis.} Bond statistics, atomic coordination, and carbon cluster size were determined from neighbor-list analysis as implemented in the Atomic Simulation Environment~(ASE) package,~\cite{Larsen-17-02} using element-specific cutoffs of \SI{1.20}{\angstrom} (\ce{C-H}), \SI{1.85}{\angstrom} (\ce{C-C}), and \SI{2.10}{\angstrom} (\ce{C-Cl}). 
The development of the carbon cluster was quantified by constructing a connectivity graph for each trajectory frame, in which carbon atoms were treated as nodes and \ce{C-C} bonds as edges. 
The size of the largest connected component was then extracted from this graph. 
Six-membered rings were identified using a minimum cycle basis algorithm on the carbon connectivity graph; 
carbon atoms belonging to these hexagonal rings were flagged. 
For visualization, non-ring atoms were rendered with increased transparency using the \texttt{Compute Property} modifiers in OVITO.~\cite{Stukowski-12-05} 

{\bf Radical analysis.} Atomic indices changed due to variations in atom count during the simulation. 
To enable time-resolved analysis, atoms were assigned persistent identifiers based on their indices in the initial frame. 
After each HCl-removal step, remaining atoms were remapped across subsequent frames using position matching under periodic boundary conditions. 
Pre-reactive-encounter times were determined by back-tracing eliminated atoms from the point of removal to the last frame in which they were covalently bonded to a carbon atom. 
A ``reactive encounter'' was defined as the first subsequent frame where the radical came within \SI{1.50}{\angstrom} of a complementary species.

In the event-based radical lifetime analysis, chlorine atoms lacking neighbors within expanded cutoffs of \SI{1.60}{\angstrom} (\ce{H-Cl}), \SI{2.30}{\angstrom} (\ce{C-Cl}), \SI{2.40}{\angstrom} (\ce{Cl-Cl}), were classified as radicals. 
Consecutive isolated frames for a given atom were merged into distinct radical events, from which formation times and durations were extracted. 
Time-resolved \ce{Cl.} radical dynamics were analyzed by correlating radical lifetimes with their respective formation times. 
A rolling mean was applied to the lifetime data to reduce noise.

For each elimination event, the parent carbon sites -- those previously bonded to the eliminated hydrogen and chlorine atoms -- were identified using the established bonding cutoffs. 
The shortest-path separation between these parent sites at the point of detachment was quantified using a topological \ce{C-C} connectivity metric, and elimination events were classified accordingly. 
For long-range or inter-chain cases, the spatial proximity was further characterized by Euclidean distances under periodic boundary conditions.

\section*{Data availability}

Data supporting this study, including the MLIP parameters and relevant structural data, will be made openly available at \url{https://github.com/vldgroup/papers-pvdc}.

\section*{Acknowledgements}

L.W. thanks L.-B.~Pașca, L.A.M.~Rosset, and N.L.~Fragapane for scientific discussions and constructive feedback on the manuscript.
L.W. acknowledges funding from the EPSRC Centre for Doctoral Training in Inorganic Chemistry for Future Manufacturing (OxICFM), EP/S023828/1. 
L.W. acknowledges the support of a Clarendon Fund Scholarship. 
This work was supported by UK Research and Innovation [grant number EP/X016188/1]. 
The authors are grateful for computational support from the UK national high performance computing service, ARCHER2, for which access was obtained via the UKCP consortium and funded by EPSRC grant ref EP/X035891/1 (see also ref.~\citenum{Beckett-24-12}).

\clearpage

\section*{References}
\vspace{2mm}

\setstretch{1}

\providecommand{\latin}[1]{#1}
\makeatletter
\providecommand{\doi}
  {\begingroup\let\do\@makeother\dospecials
  \catcode`\{=1 \catcode`\}=2 \doi@aux}
\providecommand{\doi@aux}[1]{\endgroup\texttt{#1}}
\makeatother
\providecommand*\mcitethebibliography{\thebibliography}
\csname @ifundefined\endcsname{endmcitethebibliography}  {\let\endmcitethebibliography\endthebibliography}{}

\end{document}


\emergencystretch=3em

\title{\Large {\bf Supporting Information for}\\[4mm] ``Atomistic Mechanisms of Hard Carbon Formation\\ from Polyvinylidene Chloride''}

\author[1]{Litong Wu}
\author[1]{Zitong Wu}
\author[1]{Zakariya El-Machachi}
\author[1]{Volker~L.~Deringer\thanks{volker.deringer@chem.ox.ac.uk}}

\affil[1]{Inorganic Chemistry Laboratory, Department of Chemistry, University of Oxford, Oxford, UK}

\date{}

\maketitle

\thispagestyle{empty}

\setcounter{tocdepth}{2}
{
    \setstretch{1.4}
    \tableofcontents
}

\clearpage
\setstretch{1.5}

\section{Methods}

\subsection{Dataset}

The composition of the complete dataset is summarized in Table~\ref{tab:T1}. 
For each configuration type, 10\% of the structures were reserved as the test set and used for the numerical tests in Figure~\ref{fig:S1}a,b. The remaining structures were partitioned into training and validation sets in a 9:1 ratio. 

The dataset was designed to span the chemical environments encountered during PVDC pyrolysis, from intact and decomposing polymer configurations to chlorinated molecular fragments, hydrogenated carbon networks, amorphous carbon, and graphitic configurations. 
Here, PVDC-1 denotes configurations generated using the CASTEP$+$ML approach, \cite{Stenczel-23-07} whereas PVDC-2 denotes configurations generated from iterative MLIP-driven MD.
Some configurations are taken from previous work, \cite{Rowe-20-07a, Ibragimova-25-02, Kim-25-01, Williams-25-01} as indicated in Table~\ref{tab:T1}.
Detailed descriptions of each configuration type are provided in the Dataset subsection of the Methods section in the main text. 

\begin{table}[h]
\centering
\caption{Composition of the PVDC dataset, including the number of structures, total atom counts, element-resolved atom counts, and sources for each configuration type.}
\renewcommand{\arraystretch}{1.25}
\small
\begin{tabular}{lcccccl}
\hline
\textbf{\multirow{2}{*}{Config types}} & \textbf{\multirow{2}{*}{Structures}} & \textbf{\multirow{2}{*}{Atoms}} & \multicolumn{3}{c}{\textbf{Elements}} & \textbf{\multirow{2}{*}{Sources}} \\ 
\cline{4-6} & & & \textbf{C} & \textbf{H} & \textbf{Cl} & \\ 
\hline
Isolated atoms & 3 & 3 & 1 & 1 & 1 & \\ 
\hline
PVDC-1 & 1,303 & 116,904 & 38,968 & 38,968 & 38,968 & CASTEP$+$ML~\cite{Stenczel-23-07} \\
PVDC-2 & 1,065 & 174,340 & 76,016 & 49,248 & 49,076 & Iterative MD \\ 
\hline
Amorphous C & 500 & 100,000 & 100,000 & 0 & 0 & Rowe~{\em et~al.}~\cite{Rowe-20-07a}\\
Graphite & 180 & 6,324 & 6,324 & 0 & 0 & Rowe~{\em et~al.}~\cite{Rowe-20-07a}\\
CH configurations & 509 & 66,938 & 28,930 & 38,008 & 0 & Ibragimova~{\em et~al.}~\cite{Ibragimova-25-02}\\
\hline
C/H/Cl molecules & 189 & 2,322 & 1,130 & 853 & 339 & PubChem~\cite{Kim-25-01}\\
QM9 molecules & 398 & 7,584 & 3,262 & 4,029 & 293 & QM9~\cite{Williams-25-01}\\ 
\hline
\textbf{Total} & \textbf{4,147} & \textbf{474,415} & \textbf{254,631} & \textbf{131,107} & \textbf{88,677} & \\ \hline
\label{tab:T1}
\end{tabular}
\vspace{6pt}
\raggedright
\end{table}

\clearpage

\subsection{MLIP model validation}

The MACE~\cite{Batatia-22-12} model fitted to the final dataset was validated using both global test-set errors and targeted checks relevant to the chemical and structural environments encountered during PVDC pyrolysis. 
A parity analysis assesses overall interpolation accuracy across the labeled dataset, while additional tests examine transferability to bond elongation, carbonized structures, and high-temperature dynamics.

\begin{figure}[h]
    \centering
    \includegraphics[width=\linewidth]{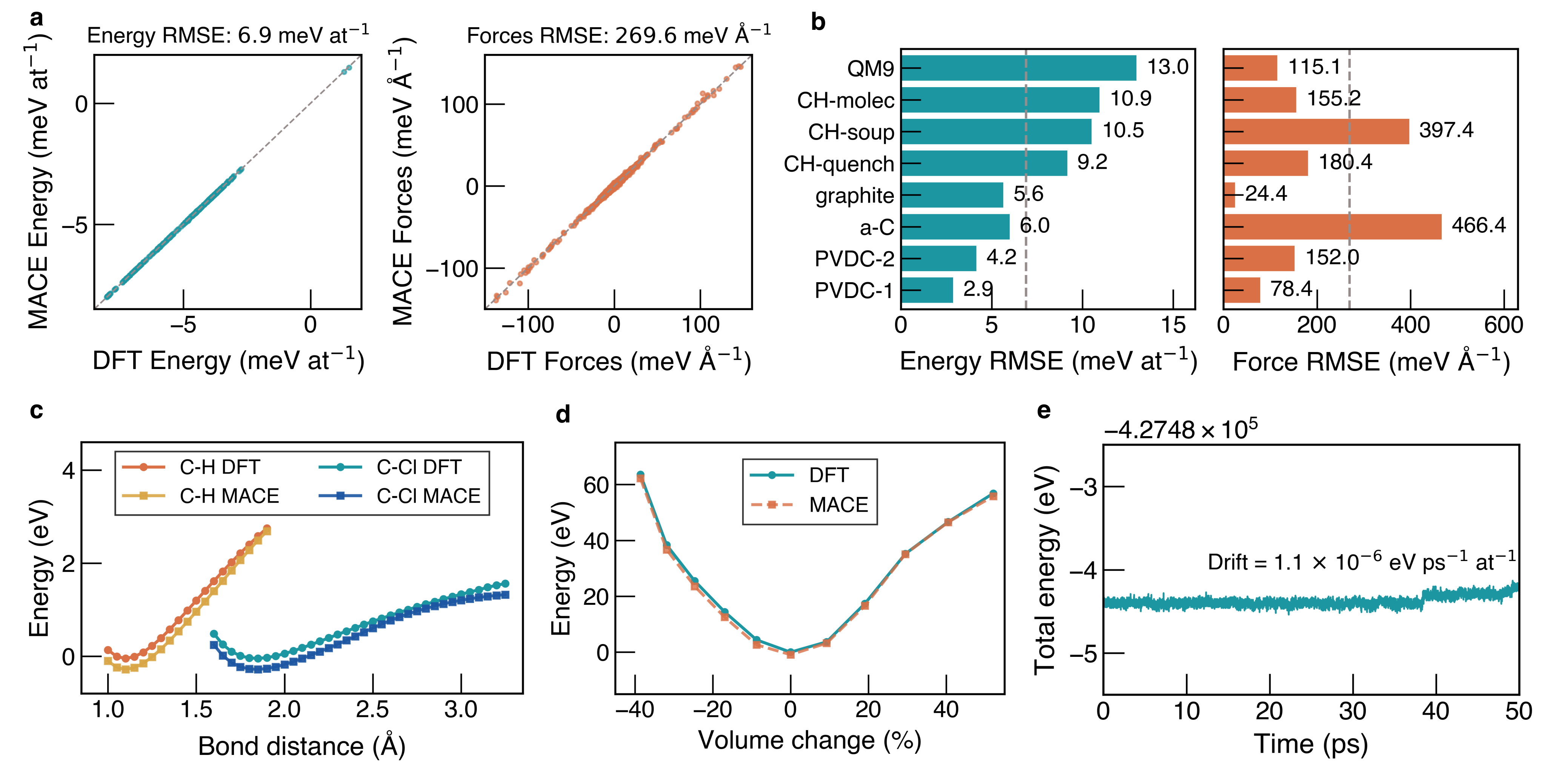}
    \caption{\textbf{Numerical validation and robustness tests of the MACE model.} 
    (\textbf{a})~Parity plots comparing MACE-predicted and DFT-computed per-atom energies~({\em left}) and forces~({\em right}) for the test set. Root-mean-square errors~(RMSEs) are indicated above each panel. 
    (\textbf{b})~Per-atom energy~({\em left}) and force~({\em right}) RMSEs resolved by configuration type. Gray dashed lines indicate the corresponding average RMSEs over the full test set.
    (\textbf{c})~MACE- and DFT-computed bond-stretching energy profiles for randomly selected \ce{C-H}~({\em orange}) and \ce{C-Cl}~({\em blue}) bonds in a single 48-atom PVDC chain. 
    The \ce{C-H} bond was stretched over \SIrange{1.0}{1.9}{\angstrom}; \ce{C-Cl} bond, \SIrange{1.6}{3.3}{\angstrom}, in increments of \SI{0.05}{\angstrom}. 
    At each sampled bond length, spin-polarized DFT geometry optimizations were performed with the selected bond length and simulation-cell dimensions held fixed.  
    Energies are reported relative to the DFT energy of the original structure.
    (\textbf{d})~Equation-of-state~(EoS) validation of a 250-atom pyrolyzed structure obtained from a small-scale MD simulation following the production protocol. 
    The simulation cell was isotropically scaled over 11 volumes spanning compressed and expanded states relative to the reference structure, after which atomic positions were optimized with the BFGS optimizer~\cite{Broyden-70, Fletcher-70-03, Goldfarb-70, Shanno-70} in ASE using the MACE model, until the maximum force was below \SI{0.01}{\electronvolt\per\angstrom}. 
    MACE energies are compared with DFT energies for the corresponding MACE-optimized structures; energies are reported relative to the original structure.
    (\textbf{e})~Microcanonical~(NVE) stability test for a 2160-atom PVDC structure initialized at \SI{3000}{\kelvin} and propagated for \SI{50}{\pico\second}.}
    \label{fig:S1}
\end{figure}

The global parity plots in Figure~\ref{fig:S1}a show that the model reproduces the DFT-labeled test set with low energy error and acceptable force error across the diverse chemical environments included in the dataset. 
The configuration-resolved errors in Figure~\ref{fig:S1}b show systematic dependence on the type of local environment sampled. 
Energy errors are relatively larger for the molecular subsets, where the smaller number of atoms per structure reduces the averaging effect present in bulk configurations. 
In contrast, the largest force errors occur for the CH ``soup'' (see Ref.~\citenum{Ibragimova-25-02}) and amorphous-carbon (see Ref.~\citenum{Rowe-20-07a} and references therein) configurations, which contain highly disordered and high-energy local environments. 
These environments produce a wider force distribution, making accurate force prediction intrinsically harder.

The constrained bond-stretching scans in Figure~\ref{fig:S1}c test the local description of \ce{C-H} and \ce{C-Cl} bond elongation, which is directly relevant to the simulated dehydrochlorination mechanism. 
The close agreement between MACE and DFT energies along these scans indicates that the model captures the energetic response to bond distortions well. 
Since only one \ce{C-H} bond and one \ce{C-Cl} bond were selected along the polymer chain, this test should be viewed as a targeted validation of bond-elongation behavior rather than a reaction-path benchmark.

The EoS comparison in Figure~\ref{fig:S1}d assesses the model's ability to describe to a dense carbonized structure representative of the later stages of pyrolysis. 
Agreement between MACE and DFT energies over compressed and expanded configurations indicates that the model remains reliable beyond PVDC-like environments and can describe the structural response of a carbon-rich network. 

The NVE test in Figure~\ref{fig:S1}e assesses the numerical stability and energy conservation of the model under high-temperature conditions. 
In an ideal NVE simulation, the total energy should remain constant throughout the trajectory; 
in practice, numerical integration errors and inaccuracies in the underlying potential-energy surface~(PES) can accumulate over time, leading to artificial energy drift. 
The very small drift of approximately \(1.1 \times 10^{-6}\,\mathrm{eV\,atom^{-1}\,ps^{-1}}\) observed over \SI{50}{\pico\second} at \SI{3000}{\kelvin} therefore indicates stable numerical integration and a smooth MACE PES under high-temperature conditions.

\clearpage

\subsection{MD sensitivity test}
\label{sec:S1.3}

To assess how sensitive the simulated pyrolysis pathway is to the density-rescaling parameter and the maximum annealing temperature, two additional simulations were performed and compared with the reference production trajectory. 
Test~1 used the same temperature schedule as the reference simulation but a larger volume decrement of \SI{33.5}{\angstrom\cubed} per removed HCl molecule, compared with \SI{30.0}{\angstrom\cubed} in the reference case. 
Test~2 used the same volume decrement as the reference simulation but a lower maximum temperature of \SI{2500}{\kelvin}. 

\begin{figure}[h]
    \centering
    \includegraphics[width=\linewidth]{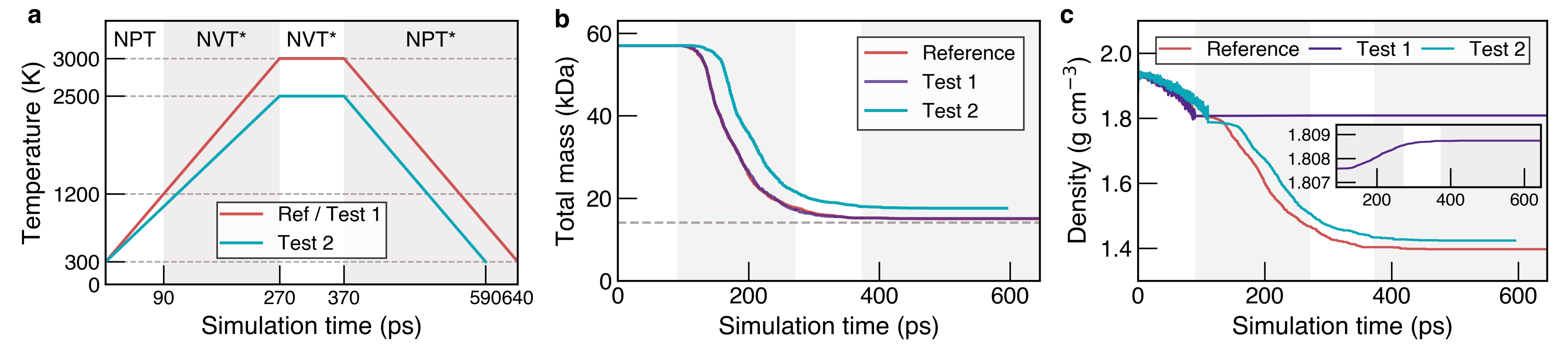}
    \caption{
    \textbf{Sensitivity of PVDC pyrolysis to simulation parameters.}
    (\textbf{a})~Temperature profiles used for the reference simulation/Test~1~({\em pink}) and Test~2~({\em blue}). 
    The reference simulation and Test~1 followed the same temperature schedule; 
    Test~2 used a lower maximum temperature of \SI{2500}{\kelvin} and a reduced heating rate to maintain the same heating duration, while the quenching rate was kept constant.
    (\textbf{b})~Cumulative mass change associated with HCl removal. 
    The gray dashed line marks the mass expected after complete HCl elimination. 
    (\textbf{c})~Density evolution for the three simulations. 
    The inset highlights the small density variation observed in Test~1 after the NPT stage.}
    \label{fig:S2}
\end{figure}

As shown in Figure~\ref{fig:S2}b, the reference simulation and Test~1 show nearly identical mass-loss profiles, indicating that the larger volume decrement has little effect on the simulated dehydrochlorination rate. 
As expected, however, Test~1 reaches a substantially higher final density as shown in Figure~\ref{fig:S2}c because each HCl-removal event is followed by a larger cell-volume reduction. 
This comparison therefore separates the chemical progress of dehydrochlorination from the density imposed by the rescaling procedure.

By contrast, lowering the maximum temperature in Test~2 delays the onset of dehydrochlorination and leads to a lower overall mass loss~(Figure~\ref{fig:S2}b), implying less complete HCl elimination. 
As seen in Figure~\ref{fig:S2}c, the density decrease is also delayed, and the final density remains slightly higher than that of the reference simulation.
This is likely because the lower extent of HCl removal leaves more mass in the simulation cell, partly offsetting the density decrease associated with cell rescaling. 
Thus, the volume decrement primarily controls the final density, whereas the maximum temperature primarily affects the extent and timing of HCl elimination.

Final-state structural metrics are summarized in Table~\ref{tab:T2}. 
Carbon coordination was computed using the same neighbor-list procedure as described in the Methods section of the main text. 
Ring statistics were computed from a carbon-only connectivity graph constructed using a \ce{C-C} cutoff of \SI{1.85}{\angstrom}. 
Rings were identified using a shortest-path ring-search algorithm~\cite{Franzblau-91-09} on the carbon connectivity graph: 
for each \ce{C-C} edge, the edge was temporarily removed and the shortest alternative path between the two bonded carbon atoms was identified; the removed edge together with this path defines the ring associated with that edge.
Ring abundances are defined as the percentage of unique carbon atoms participating in at least one \(n\)-membered ring. 
Because an atom may belong to multiple rings in fused networks, the percentages are not mutually exclusive, and should be interpreted as internally consistent comparative metrics rather than unique topological decompositions.

\begin{table}[h]
\centering
\caption{Final-state properties from the MD sensitivity tests.}
\renewcommand{\arraystretch}{1.25}
\small
\begin{tabular}{llccc}
\hline
\multicolumn{2}{l}{} & \textbf{Reference} & \textbf{Test~1} & \textbf{Test~2} \\
\hline
\multicolumn{2}{l}{$T_\mathrm{max}$ (K)} & 3000 & 3000 & 2500 \\
\multicolumn{2}{l}{$\Delta V_{\ce{HCl}}$ (\AA$^3$)} & 30.0 & 33.5 & 30.0 \\
\hline
\multicolumn{2}{l}{HCl elimination (\%)} & 97.9 & 97.7 & 90.6 \\
\multicolumn{2}{l}{Final density (g\,cm$^{-3}$)} & 1.40 & 1.81 & 1.42 \\
\hline
\multirow{3}{*}{\begin{tabular}[c]{@{}l@{}}C coordination\\number (\%)\end{tabular}} 
& 2-fold & 4.8 & 1.1 & 5.7 \\
& 3-fold & 94.6 & 97.4 & 92.9 \\
& 4-fold & 0.6 & 1.4 & 1.4 \\
\hline
\multirow{3}{*}{\begin{tabular}[c]{@{}l@{}}C ring\\abundance (\%)\end{tabular}} 
& 5-membered & 50.9 & 57.2 & 51.3 \\
& 6-membered & 80.4 & 81.2 & 70.7 \\
& 7-membered & 25.3 & 38.0 & 24.5 \\
\hline
\end{tabular}
\label{tab:T2}
\end{table}

All three simulations produce predominantly threefold-coordinated carbon networks, despite differences in final density and percentage HCl elimination. 
Increasing the volume decrement slightly increases the fraction of threefold-coordinated carbon, consistent with the higher final density favoring a more connected carbon network.

The ring statistics further indicate that non-hexagonal motifs are retained across all tested protocols. 
Although the precise distribution of ring sizes varies, five- and seven-membered rings remain present in all final structures. 
This supports the conclusion that odd-membered-ring formation is not an artefact of the reference density-rescaling parameter or maximum temperature.

\clearpage

\section{Results}

\subsection{Molecular product statistics}

To quantify the relative abundance of molecular products formed during the production trajectory, we identified \ce{H2}, HCl, and \ce{Cl2} molecules using distance-based bonding criteria and a minimum lifetime threshold of \SI{0.01}{\pico\second}. 
The bonding cutoffs, listed in Table~\ref{tab:T3}, slightly exceed the corresponding gas-phase equilibrium bond lengths (\SI{0.741}{\angstrom}, \SI{1.275}{\angstrom}, and \SI{1.988}{\angstrom}, respectively), while remaining sufficiently restrictive to avoid counting non-bonded encounters as molecules.~\cite{Huber-79}

\begin{table}[h]
\centering
\caption{Molecular products identified during the production trajectory. Fractions are reported relative to the total number of identified \ce{H2}, HCl, and \ce{Cl2} molecular events. Molecules were identified using the distance-based bonding cutoffs listed in the second column. HCl lifetime statistics are influenced by the periodic HCl-removal protocol and are therefore not reported.}
\renewcommand{\arraystretch}{1.25}
\small
\begin{tabular}{cc|cc|ccc}
\hline
\multirow{2}{*}{\textbf{Molecule}} &
\multirow{2}{*}{\textbf{Cutoff (\AA)}} &
\multirow{2}{*}{\textbf{Count}} &
\multirow{2}{*}{\textbf{Fraction (\%)}} &
\multicolumn{3}{c}{\textbf{Lifetime (ps)}} \\
\cline{5-7}
& & & & \textbf{Mean} & \textbf{Minimum} & \textbf{Maximum} \\
\hline
\ce{H2}  & 0.90 & 13   & 0.5  & 1.93 & 0.01 & 24.99 \\
HCl      & 1.30 & 1973 & 83.1 & --   & --   & --    \\
\ce{Cl2} & 2.20 & 389  & 16.4 & 0.35 & 0.01 & 7.54  \\
\hline
\textbf{Total} & -- & \textbf{2375} & \textbf{100.0} & -- & -- & -- \\
\hline
\end{tabular}
\label{tab:T3}
\end{table}

The statistics confirm that HCl is the dominant molecular product in the simulated decomposition pathway, accounting for 83.1\% of all identified molecular events. 
\ce{Cl2} formation is observed more frequently than \ce{H2}, but \ce{Cl2} remains a secondary product with relatively low lifetime. 
The dominance of HCl supports the use of HCl removal as the principal volatile-loss channel in the MD protocol and is consistent with experimental studies identifying HCl as the dominant volatile species evolved during PVDC degradation.

\clearpage

\subsection{Kinetic analysis of dehydrochlorination}

\begin{figure}[h]
    \centering
    \includegraphics[width=0.45\linewidth]{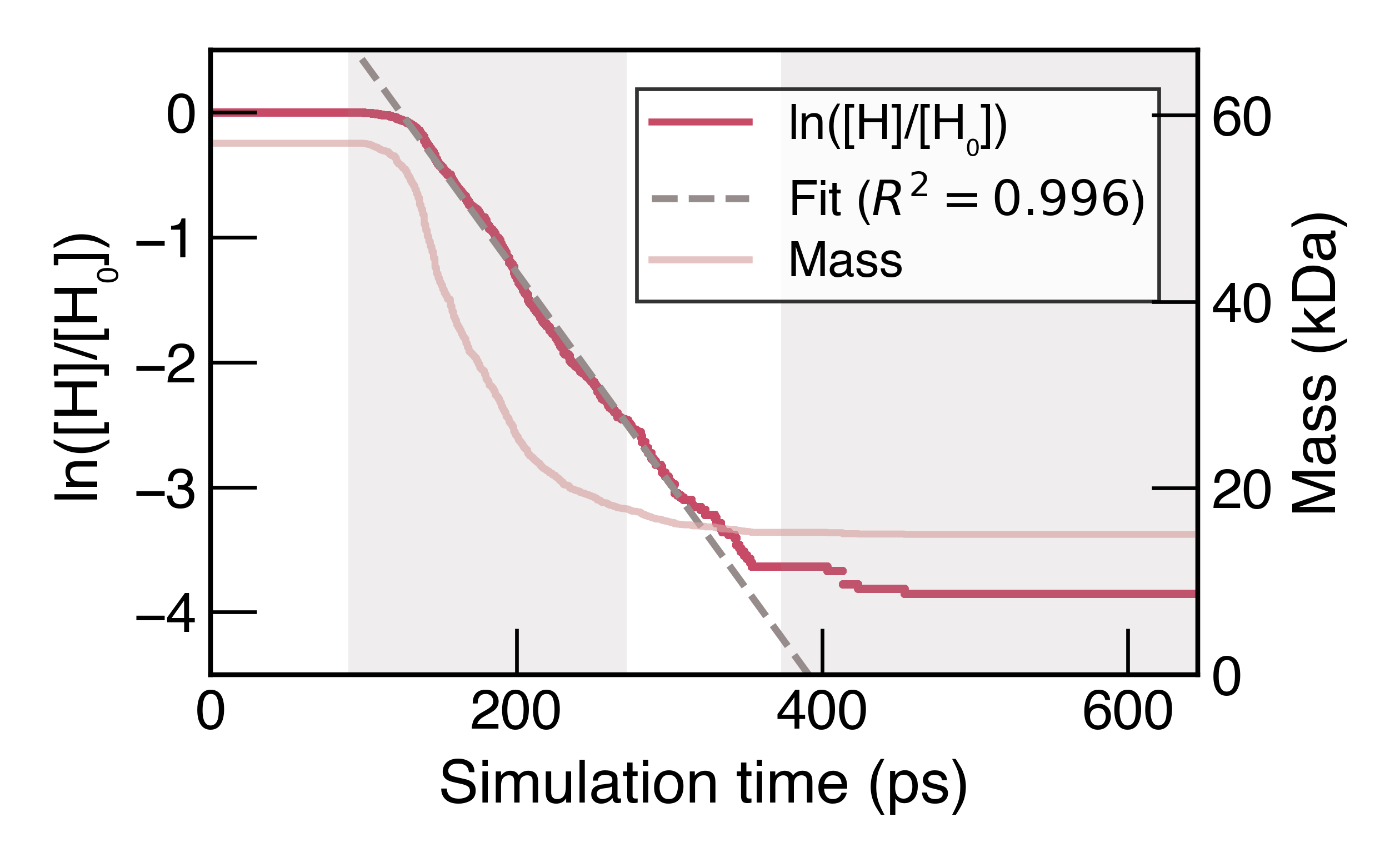}
    \caption{\textbf{First-order kinetic analysis of simulated PVDC dehydrochlorination.}
    Logarithmic decay of the hydrogen content in the simulation cell, used as a proxy for the remaining eliminable sites, is shown in burgundy. 
    The gray dashed line shows a linear fit over \SIrange{130}{320}{\pico\second}~($R^{2}=0.996$), and the translucent pink curve shows the cumulative mass change associated with HCl removal.}
    \label{fig:S3}
\end{figure}

Since dehydrochlorination depletes hydrogen and chlorine stoichiometrically, the remaining hydrogen content provides a convenient proxy for the progress of dehydrochlorination. 
The main dehydrochlorination regime exhibits a nearly linear logarithmic decay of the reactant, supporting the use of an apparent pseudo-first-order description in the main text. 
The fitting window was selected to provide a linear regime, excluding both the initial induction period before sustained HCl evolution and the late-stage regime in which the remaining H/Cl-containing sites become depleted. 
Although the window was chosen empirically, it spans a substantial portion of the dehydrochlorination process. 
Nonetheless, this analysis should be interpreted as an empirical description of the simulated trajectory rather than as an experimental rate law, because the production protocol is accelerated and non-isothermal.

\clearpage

\subsection{Diffusivity analysis}

\begin{figure}[h]
    \centering
    \includegraphics[width=0.42\linewidth]{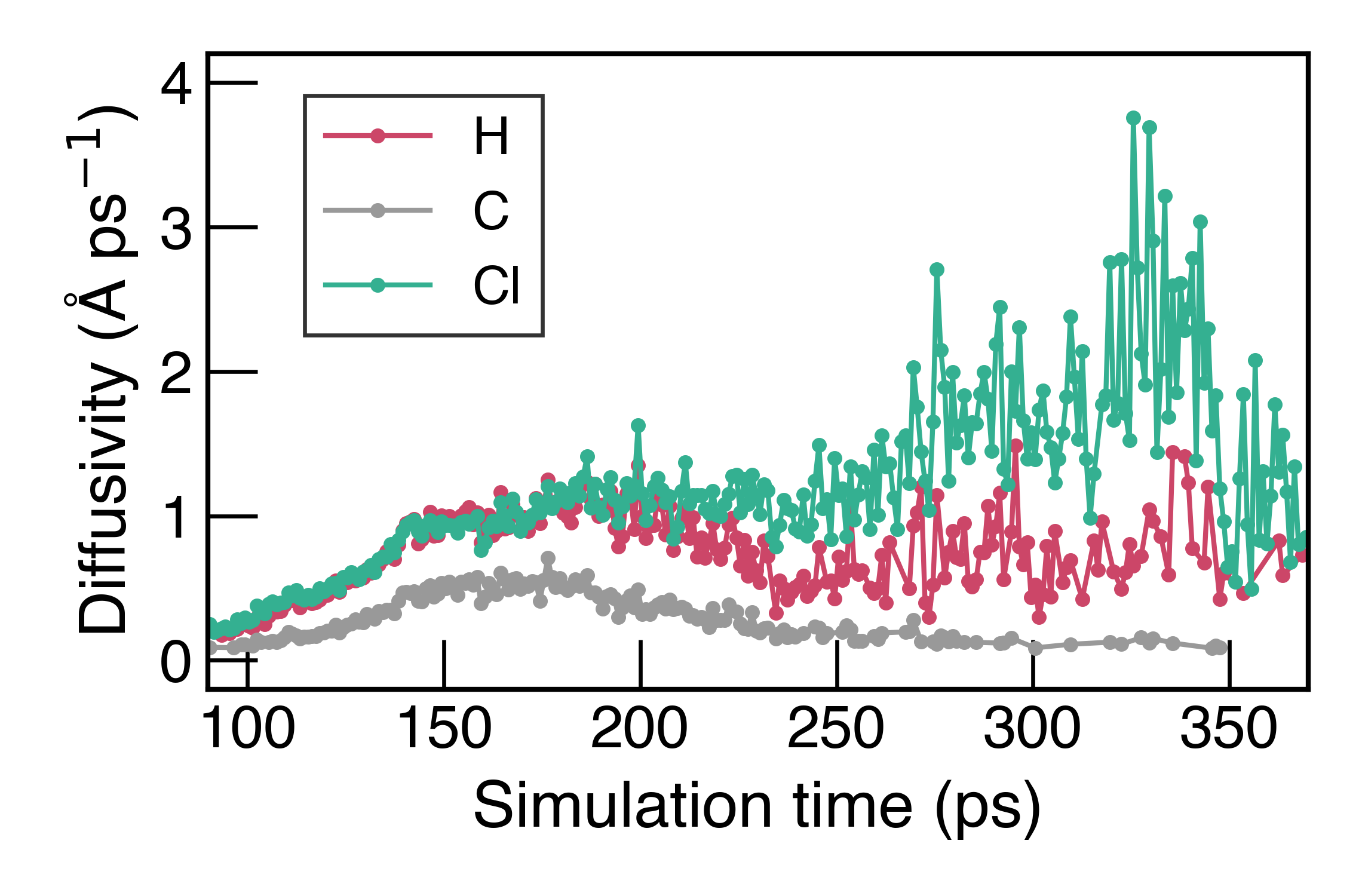}
    \caption{\textbf{Segment-wise apparent mobility of carbon, hydrogen, and chlorine during PVDC pyrolysis.}
    The apparent diffusivity of different elemental species during the NVT$^\ast$ stages was estimated from the slope of the mean-squared displacement~(MSD) within each MD segment, using only segments with approximately linear MSD growth (\(R^{2}>0.7\)). 
    Each point corresponds to the mid-point of an independent simulation segment. 
    Since the MSD computation is reset at the start of each segment,~\cite{Thompson-22} the values should be interpreted as segment-wise apparent mobilities rather than long-time diffusion coefficients.}
    \label{fig:S4}
\end{figure}

The segment-wise apparent mobility of carbon shows a modest rise during the period of active clustering, followed by a gradual decline as structural rigidity develops. 
Hydrogen remains more mobile than carbon but does not display the pronounced late-stage enhancement seen for chlorine. 
In contrast, the apparent diffusivity of chlorine is generally higher than that of hydrogen and much higher than that of carbon; 
although this metric includes both bonded and detached chlorine atoms, it is consistent with the idea that chlorine-containing species act as the primary mobile participants in HCl-forming events.
After 250 ps, the chlorine diffusivity fluctuates strongly, with several high-mobility spikes. 
This behavior suggests that late-stage dehydrochlorination is not uniform: some segments contain highly mobile chlorine species, while others exhibit more restricted chlorine mobility, consistent with a regime in which HCl formation becomes increasingly encounter-limited as reactive H-containing sites are depleted.

We note that the LAMMPS \texttt{compute msd} output reflects the mobility of all atoms in the specified group. 
Accordingly, the resulting curves should not be interpreted as direct comparisons between \ce{Cl.} radicals and isolated hydrogen species. 
In particular, the high hydrogen mobility partly reflects the inherently larger thermal velocity of hydrogen and the presence of highly mobile gaseous HCl molecules that have formed but not yet been removed.

\clearpage

\subsection{Evolution of local coordination environments}
\label{sec:S2.4}

To resolve how dehydrochlorination changes the immediate bonding environment of carbon atoms, we classified under-coordinated carbon sites according to their bonded neighbors. 

\begin{figure}[h]
    \centering
    \includegraphics[width=\linewidth]{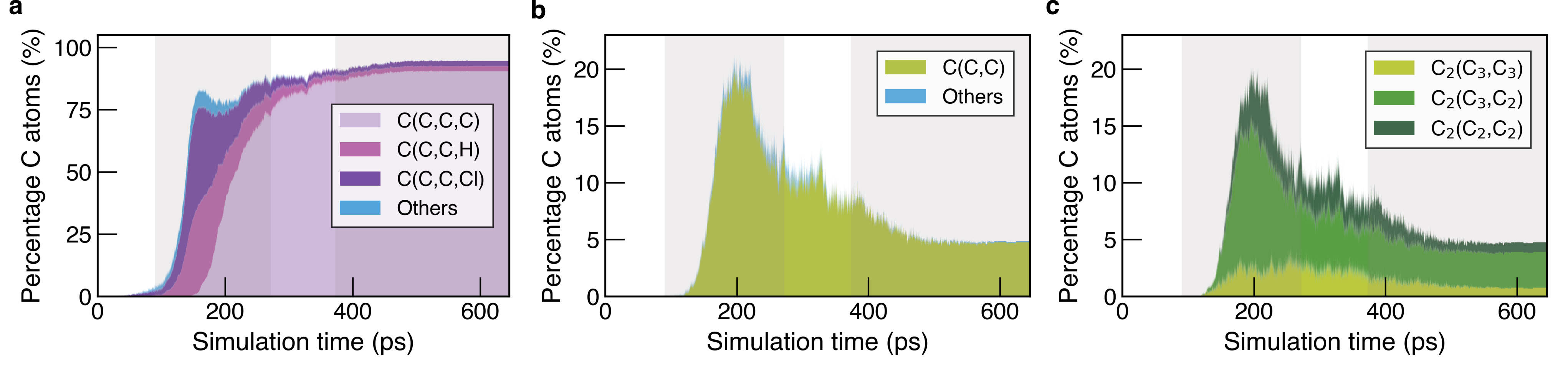}
    \caption{\textbf{Temporal evolution of local coordination environments of carbon atoms.}
    (\textbf{a})~Fractions of threefold-coordinated carbon atoms classified by neighbor environments. 
    For example, C(C,C,Cl) denotes a threefold-coordinated carbon atom bonded to two carbon atoms and one chlorine atom. 
    (\textbf{b})~Fractions of twofold-coordinated carbon atoms classified by neighbor environments. 
    (\textbf{c})~Fractions of twofold-coordinated carbon atoms further classified by the coordination states of their carbon neighbors. 
    For example, C$_2$(C$_3$,C$_2$) denotes a twofold-coordinated carbon atom bonded to one threefold-coordinated carbon atom and one twofold-coordinated carbon atom.
    All percentages are normalized by the total number of carbon atoms in the simulation cell.}
    \label{fig:S5}
\end{figure}

In Figure~\ref{fig:S5}a, the very early stage of the reaction is dominated by C(C,C,Cl) environments, consistent with the initial loss of \ce{C-Cl} bonds from fourfold-coordinated PVDC carbon atoms. 
As dehydrochlorination progresses, C(C,C,H) environments emerge, reflecting subsequent \ce{C-H} bond cleavage. 
A minor population of other threefold-coordinated environments, primarily C(C,H,H) and C(C,Cl,Cl), is also observed; these species are attributed to transient \ce{C-C} bond scission under high-temperature conditions. 

At later stages, the growth of C(C,C,C) environments is accompanied by decreasing populations of both C(C,C,Cl) and C(C,C,H), indicating continued dehydrochlorination from already unsaturated threefold-coordinated carbon sites. 
This interpretation is supported by the concurrent increase in C(C,C) environments in Figure~\ref{fig:S5}b. 
The subsequent decrease in C(C,C) environments, together with the growth of C(C,C,C) environments, suggests that these under-coordinated carbon sites are rapidly consumed through \ce{C-C} cross-linking.

For twofold-coordinated carbon atoms, classification by immediate neighbors is less diagnostic because nearly all of them are bonded to two carbon atoms. 
Further classification according to the coordination states of neighboring carbon atoms distinguishes C$_2$(C$_3$,C$_3$) network bridges, C$_2$(C$_3$,C$_2$) chain--network junctions, and C$_2$(C$_2$,C$_2$) chain-segment sites. 
The dominance of C$_2$(C$_3$,C$_2$) environments indicates that twofold-coordinated carbon atoms occur more often at chain--network junctions than in extended chain segments; 
in other words, longer cumulene-like chains are disfavored. 
The distinct behavior of C$_2$(C$_3$,C$_3$) sites is analyzed further in Section~\ref{sec:S2.7}, where their geometry, energetic signature, and lifetime are compared with other carbon local environments.

\clearpage

\subsection{Topological evolution}

To establish the link between local topology and curvature in the developing carbon network, we analyze the evolution of ring populations alongside a local measure of non-planarity for threefold-coordinated carbon atoms using the out-of-plane~(OOP) distance.

\begin{figure}[h]
    \centering
    \includegraphics[width=\linewidth]{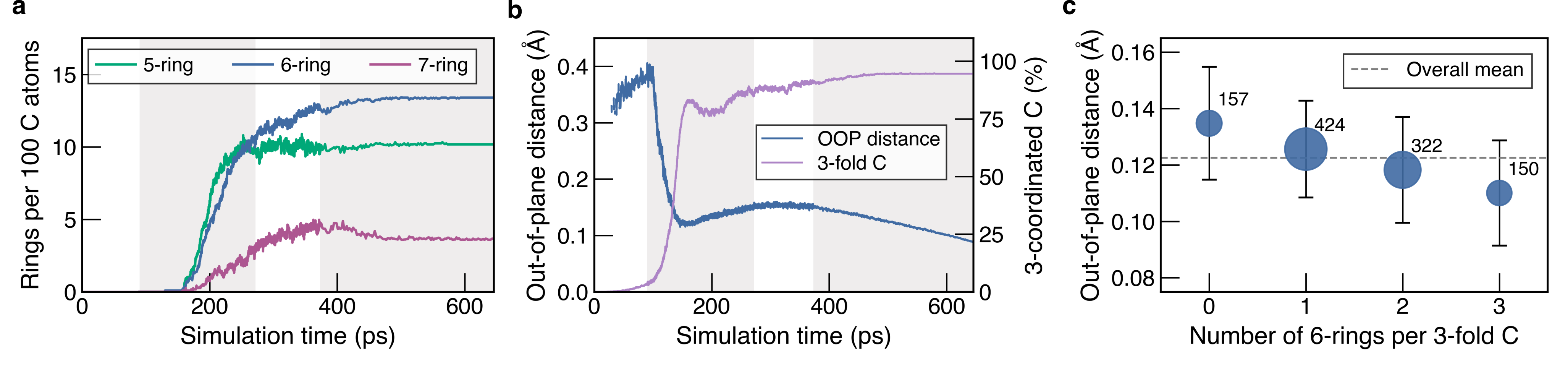}
    \caption{\textbf{Topological analysis of the carbon network.}
    (\textbf{a})~Evolution of the number of five-, six-, and seven-membered rings per 100 carbon atoms, computed using the ring-identification procedure described in Section~\ref{sec:S1.3}.
    (\textbf{b})~Evolution of the average OOP distance of threefold-coordinated carbon atoms ({\em blue}). 
    Evolution of the percentage of threefold-coordinated carbon atoms is shown in purple for reference.
    (\textbf{c})~Average OOP distance of threefold-coordinated carbon atoms during the final \SI{170}{\pico\second} of quenching ({\SIrange{2000}{300}{\kelvin}}), where the ring counts have stabilized. 
    Carbon atoms are categorized by the number of six-membered rings to which they belong. 
    Whiskers indicate standard deviations, labels indicate the average number of carbon atoms belonging to each category per frame, and are proportional to the marker size. 
    The dashed line marks the average OOP distance across all threefold-coordinated carbon atoms in the same time window.}
    \label{fig:S6}
\end{figure}

As shown in Figure~\ref{fig:S6}a, carbon rings begin to form rapidly following the onset of carbon clustering. 
Five-membered rings appear slightly earlier and initially increase more rapidly in number than six-membered rings. 
In contrast, six-membered rings continue to accumulate over a longer period and ultimately become the most abundant ring motif in the final network. 
Seven-membered rings emerge more gradually and remain less abundant overall, but nevertheless constitute a persistent population of non-hexagonal defects throughout the later stages of pyrolysis. 
Interestingly, the slight decrease in the seven-membered-ring population at the start of quenching indicates that a fraction of non-hexagonal motifs remains structurally dynamic and can be annealed into more stable motifs through local rearrangement.

For each threefold-coordinated carbon atom, the OOP distance (shown in blue in Figure~\ref{fig:S6}b) was defined as the perpendicular distance of the central atom from the plane formed by its three bonded carbon neighbors. 
Before sustained dehydrochlorination, transient \ce{C-Cl} bond cleavage can generate a small amount of formally threefold-coordinated carbon atoms that still retain the local tetrahedral geometry of the precursor polymer, leading to relatively large OOP values. 
As dehydrochlorination progresses and threefold-coordinated carbon environments become increasingly prevalent, the average OOP distance decreases rapidly. 
A modest increase during the high-temperature stage is attributed to thermal distortion of the developing carbon network. 
During quenching, the OOP distance decreases continuously and stabilizes at approximately \SI{0.1}{\angstrom}, indicating indicating partial relaxation toward locally planar sp$^2$-like environments under the constraints of the cross-linked network. 

To assess the relationship between local ring topology and non-planarity, threefold-coordinated carbon atoms from the final \SI{170}{\pico\second} of the trajectory were grouped according to the number of six-membered rings in which they participate, and the corresponding OOP-distance distributions are shown in Figure~\ref{fig:S6}c. 
A clear trend is observed: carbon atoms associated with a larger number of six-membered rings exhibit, on average, smaller OOP distances and therefore greater local planarity. 
This correlation supports the interpretation that incomplete hexagonal ordering contributes to local non-planarity and therefore curvature of the developing hard-carbon network.

\clearpage

\subsection{Atomic-energy descriptors of different local environments}

To further characterize the relative stability of different carbon environments, we analyzed the MACE-predicted per-atom energy contributions for carbon atoms with different local connectivity. 
Although atomic-energy decompositions in MLIPs are model-dependent and are not unique quantum-mechanical observables,~\cite{Chong-2023} they can provide useful comparative descriptors when interpreted within a fixed model. 
Here, we use these values to compare local carbon environments across the trajectory and to examine how local energetic trends correlate with ring topology and OOP distortion in the final carbon network.

\begin{figure}[h]
    \centering
    \includegraphics[width=\linewidth]{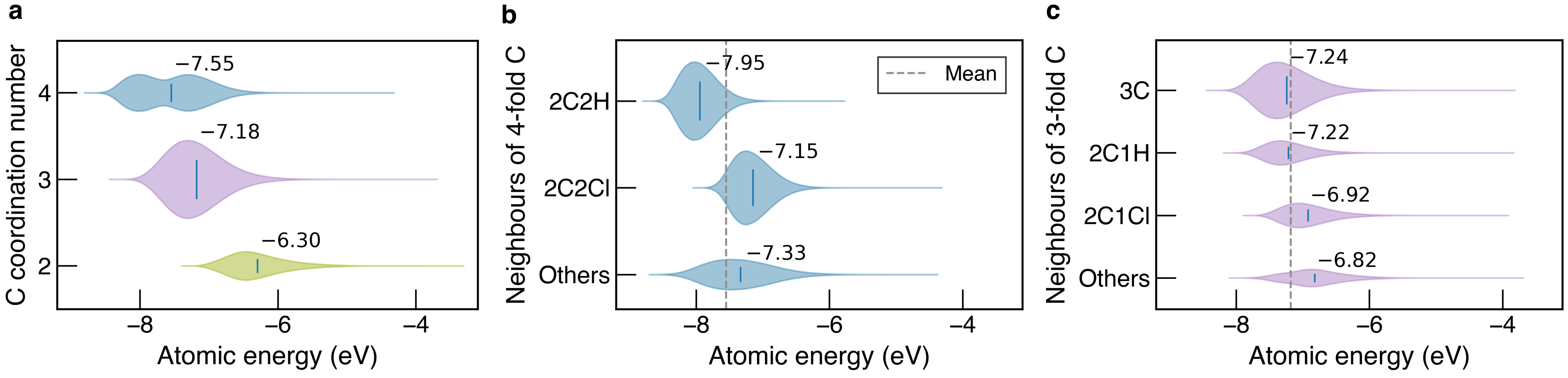}
    \caption{\textbf{MACE-predicted atomic-energy distributions for carbon atoms in different local environments.}
    (\textbf{a})~Atomic-energy distributions for carbon atoms categorized by coordination number across all simulation frames. 
    Distributions are shown for fourfold-, threefold-, and twofold-coordinated carbon atoms; 
    the blue line within each distribution indicates the corresponding mean value, and the distribution areas are proportional to the respective populations. 
    (\textbf{b})~Atomic-energy distributions for fourfold-coordinated carbon atoms categorized by immediate neighbor environment: C(C,C,H,H), C(C,C,Cl,Cl), and other fourfold-coordinated environments. 
    The gray dashed line indicates the mean atomic energy of all fourfold-coordinated carbon atoms. 
    (\textbf{c})~Atomic-energy distributions for threefold-coordinated carbon atoms categorized as C(C,C,C), C(C,C,H), C(C,C,Cl), and other threefold-coordinated environments.}
    \label{fig:S7}
\end{figure}

Fourfold-coordinated carbon atoms exhibit the lowest average MACE atomic-energy contributions among the coordination states considered. 
This likely reflects their intrinsic stability as well as their predominance in the intact PVDC precursor during the low-temperature stage of the simulation. 
Within this class (Figure~\ref{fig:S7}b), C(C,C,Cl,Cl) environments are systematically higher in energy than C(C,C,H,H) environments, consistent with the greater susceptibility of \ce{C-Cl} bonds to activation and cleavage during dehydrochlorination. 
The remaining fourfold-coordinated environments show broader and generally higher-energy distributions, as they were primarily generated through radical recombination during early-stage cross-linking processes.

Although threefold-coordinated carbon atoms exhibit higher average atomic-energy contributions than fourfold-coordinated carbon atoms when sampled over the full trajectory, this comparison should be interpreted with care as threefold-coordinated environments are populated during the high-temperature dehydrochlorination and carbonization stages. 
Within the threefold-coordinated class (Figure~\ref{fig:S7}c), C(C,C,C) environments are the lowest in energy, followed by C(C,C,H) and C(C,C,Cl), consistent with the stabilization of carbon atoms as they become incorporated into the carbon network. 
Twofold-coordinated carbon atoms are not further classified by immediate neighbor identity, because the vast majority are bonded to two carbon atoms; their secondary-neighbor environments are analyzed separately in Section~\ref{sec:S2.7}.

\begin{figure}[h]
    \centering
    \includegraphics[width=\linewidth]{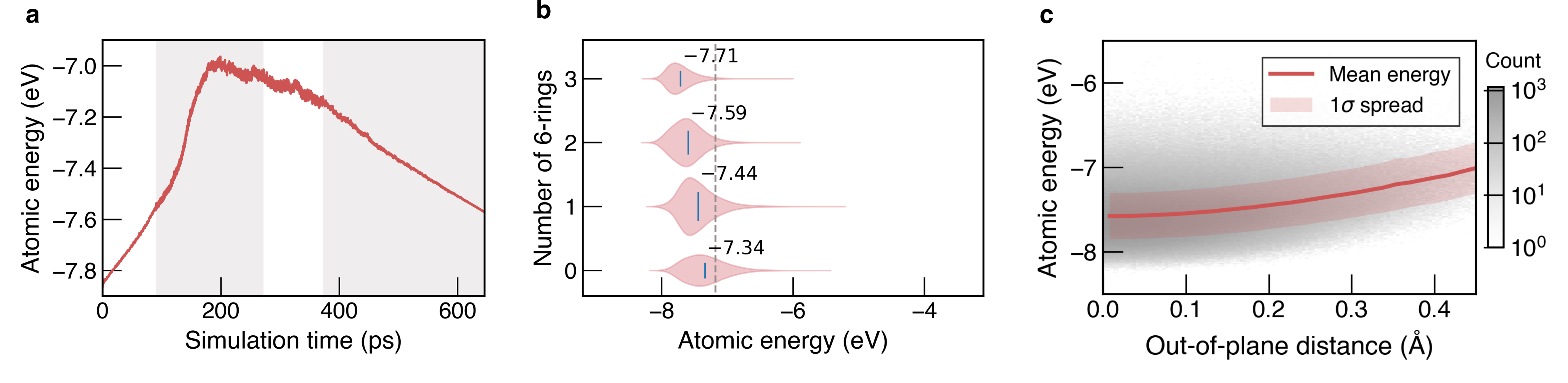}
    \caption{\textbf{Temporal evolution of MACE atomic energies and their correlation with local topology.}
    (\textbf{a})~Evolution of the frame-averaged MACE atomic-energy contributions of carbon atoms. 
    (\textbf{b})~MACE atomic-energy distributions of threefold-coordinated carbon atoms from the final \SI{170}{\pico\second} of the trajectory, categorized by the number of six-membered rings in which each atom participates. 
    (\textbf{c})~Correlation between MACE atomic-energy contribution and OOP distance for threefold-coordinated carbon atoms from the final \SI{170}{\pico\second}. 
    The heat map shows the occurrence of individual carbon environments, with denser colors reflecting higher abundance. The pink line shows the mean atomic energy as a function of OOP distance; the shaded region denotes one standard deviation.}
    \label{fig:S8}
\end{figure}

The mean carbon atomic energy initially increases with temperature, reflecting thermal bond activation in the precursor together with the progressive conversion of fourfold-coordinated carbon atoms into threefold-coordinated environments. 
The emergence of twofold-coordinated carbon atoms (around \SI{160}{\pico\second}; Figure~3 of the main manuscript) coincides with a second sharp increase in the average atomic energy, consistent with their role as energetically unfavorable intermediates. 
At later stages, the average energy decreases steadily as twofold-coordinated sites are consumed through \ce{C-C} cross-linking and the carbon network rearranges toward more stable threefold-coordinated configurations.

Figures~\ref{fig:S8}b,c show a clear correlation between ring topology, non-planarity, and MACE atomic-energy contributions during the \SIrange{2000}{300}{\kelvin} quenching stage, where the ring populations have largely stabilized. 
Threefold-coordinated carbon atoms participating in a larger number of six-membered rings exhibit both lower average atomic-energy contributions and lower maximum atomic energies. 
Similarly, the increasingly positive slope of the mean-energy curve in Figures~\ref{fig:S8}c suggests that threefold-coordinated carbon atoms with larger OOP distances are associated with increasingly higher-energy local environments. 
Together, these trends indicate that hexagon-rich, locally planar motifs are energetically more favorable within the MACE energy decomposition, whereas incompletely hexagonal topologies are associated with enhanced curvature and higher local energetic cost.

\clearpage

\subsection{Structural signatures of under-coordinated carbon intermediates}
\label{sec:S2.7}

To further characterize the local structures during pyrolysis, we analyzed the evolution of \ce{C-C} bond lengths, C-centered bond angles, and MACE-predicted atomic-energy contributions for carbon atoms with different coordination numbers. 
As twofold-coordinated carbon atoms show the largest deviations from idealized local geometries, we further resolved these environments according to the coordination states of their neighboring carbon atoms, as defined in Section~\ref{sec:S2.4}.

\begin{figure}[h]
    \centering
    \includegraphics[width=\linewidth]{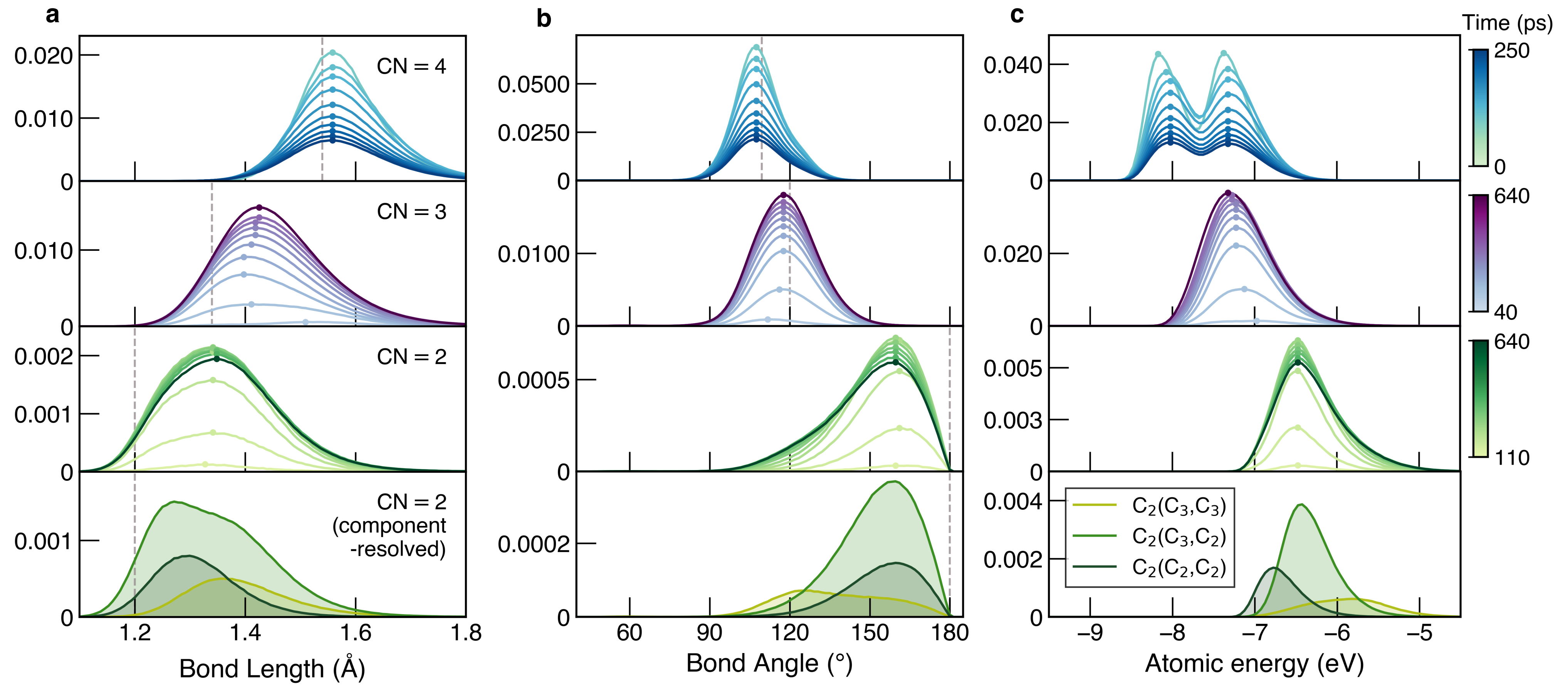}
    \caption{\textbf{Geometric and energetic signatures of local carbon environments during PVDC pyrolysis.}
    Time-resolved cumulative distributions of local descriptors: 
    (\textbf{a})~\ce{C-C} bond lengths, 
    (\textbf{b})~C-centered bond angles, and 
    (\textbf{c})~MACE-predicted atomic-energy contributions. 
    In each column, the upper three panels show distributions grouped by carbon coordination number. 
    Curves correspond to cumulative kernel-density estimates, normalized by the number of contributing atoms accumulated up to each time interval, and are colored according to simulation time. 
    Different time windows, indicated by the corresponding color bars on the right, are used for different coordination states to focus on the periods in which those environments are appreciably populated: \SIrange{0}{250}{\pico\second} for fourfold-coordinated carbon, \SIrange{40}{640}{\pico\second} for twofold-coordinated carbon, and \SIrange{110}{640}{\pico\second} for threefold-coordinated carbon. 
    Scatter points mark the evolving maxima of the corresponding distributions. 
    The lower panels resolve the same descriptors for twofold-coordinated carbon atoms according to secondary-neighbor connectivity, without time resolution for better visual clarity. 
    Gray dashed lines in panels~(\textbf{a}) and~(\textbf{b}) indicate reference values for the \ce{C_{sp$^3$}-C_{sp$^3$}} (ethane), \ce{C_{sp$^2$}-C_{sp$^2$}} (ethylene), and \ce{C_{sp}-C_{sp}} (acetylene) bond lengths;~\cite{Allen-87} and the tetrahedral, trigonal-planar, and linear C-centered bond angles, respectively.}
    \label{fig:S9}
\end{figure}

In Figure~\ref{fig:S9}a, the \ce{C-C} bond lengths between fourfold-coordinated carbon atoms remain close to typical \ce{C_{sp$^3$}-C_{sp$^3$}} bond lengths during the early stage of the simulation, while their population decreases as dehydrochlorination proceeds. 
Threefold-coordinated carbon atoms become increasingly populated during carbonization, and their characteristic \ce{C-C} bond length approaches approximately \SI{1.42}{\angstrom} toward the end of the trajectory, close to \ce{C_{sp$^2$}-C_{sp$^2$}} bond lengths observed in graphite. 
The small shift to longer bond lengths for both $\text{CN} = 4$ and $\text{CN} = 3$ may reflect residual tensile strain or the relatively open hard-carbon network formed under the density-calibrated protocol.
The bond-length distribution of twofold-coordinated carbon remains broader despite lower population, and noticeably shifted relative to the idealized sp-bonded reference, indicating that these sites are not simply relaxed linear sp-carbon motifs, but strained intermediates embedded within the carbon network.

The corresponding bond-angle distributions in Figure~\ref{fig:S9}b support this interpretation. 
Fourfold-coordinated carbon atoms remain close to the tetrahedral reference angle, with minimal variation with simulation time. 
At early times, some threefold-coordinated carbon atoms retain bond angles closer to tetrahedral geometry, consistent with local configurations that have lost a substituent but have not yet fully relaxed toward trigonal-planar geometry~(Figure~\ref{fig:S6}b). 
As the network matures, the threefold-coordinated population shifts toward angles slightly below \(120^\circ\), consistent with locally sp$^2$-like carbon environments distorted by curvature and topological disorder. 
In contrast, twofold-coordinated carbon atoms show the largest deviation from ideal linear geometry, indicating that these sites are strongly bent and constrained by the surrounding network.

The MACE atomic-energy distributions in Figure~\ref{fig:S9}c provide a complementary model-based descriptor of local stability. 
Fourfold-coordinated carbon atoms show distinct bimodal populations, consistent with the different energetic environments of C(C,C,H,H) and C(C,C,Cl,Cl) motifs discussed in Figure~\ref{fig:S7}. 
As the temperature increases, both populations shift to higher energy, reflecting thermal bond activation. 
Threefold-coordinated carbon atoms initially sample relatively high-energy environments, because early threefold-coordinated species are dominated by C(C,C,Cl)-type motifs generated by \ce{C-Cl} cleavage. 
At later times, the distribution shifts to lower energies as more stable C(C,C,C) environments form and the network relaxes during quenching. 

The anomalous geometry and high energy of twofold-coordinated carbon atoms motivated a more detailed classification by secondary-neighbor connectivity in the last panel of each column in Figure~\ref{fig:S9}. 
Among these environments, C$_2$(C$_2$,C$_2$) sites show the shortest \ce{C-C} bonds, consistent with greater cumulene-like s-character in the bonding. 
C$_2$(C$_3$,C$_2$) environments display a broader, bimodal bond-length distribution, reflecting the asymmetry between the bond to a twofold-coordinated neighbor and the bond to a threefold-coordinated network site. 
C$_2$(C$_3$,C$_3$) environments show the longest \ce{C-C} bonds. 
Moreover, in contrast to the other two C$_2$ environments, whose bond-angle distributions peak near \(160^\circ\), the bond-angle distribution of C$_2$(C$_3$,C$_3$) sites is centered much closer to trigonal values.
This suggests that these C$_2$(C$_3$,C$_3$) atoms are geometrically more similar to distorted threefold-coordinated network sites, but lack one formal \ce{C-C} connection under the adopted bonding criterion .

This interpretation is reinforced by the atomic-energy analysis. 
C$_2$(C$_3$,C$_3$) sites have the highest MACE atomic-energy contributions among the twofold-coordinated environments and are substantially higher in energy than typical threefold-coordinated carbon atoms, indicating that they are not simply misclassified threefold-coordinated carbon sites arising from a marginal cutoff choice. 
Instead, they appear to be intermediates during high-temperature network rearrangement. 

\begin{figure}[h]
    \centering
    \includegraphics[width=\linewidth]{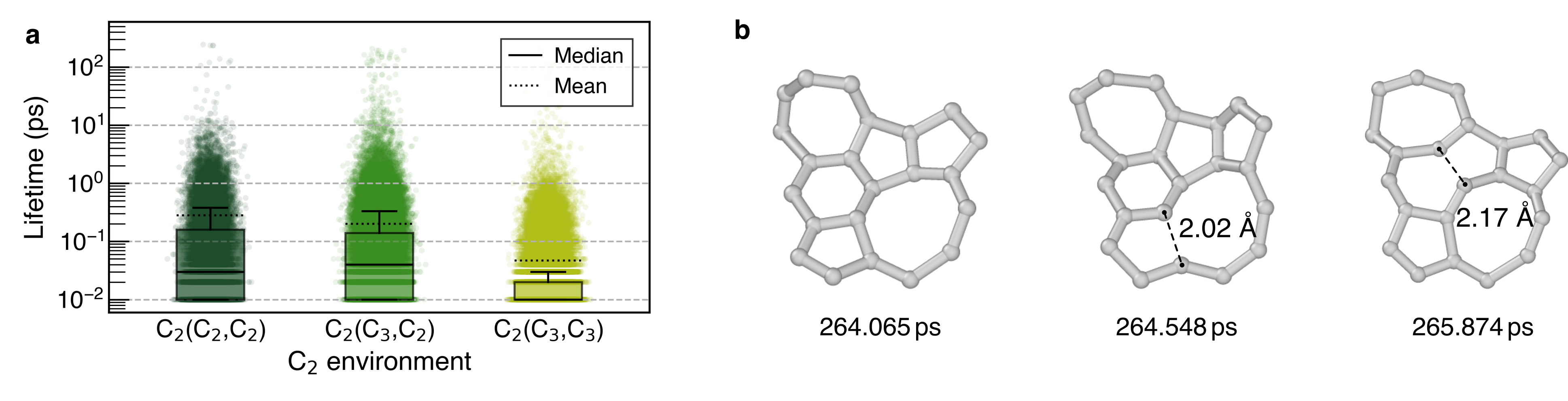}
    \caption{\textbf{Lifetimes of under-coordinated carbon intermediates}
    (\textbf{a})~Lifetime distributions of twofold-coordinated carbon atoms classified by secondary-neighbor connectivity. 
    Boxes span the interquartile range; solid and dotted lines denote the median and mean, respectively; whiskers extend to \(1.5\times\) the interquartile range.
    (\textbf{b})~Representative fragments illustrating the transient appearance and disappearance of C$_2$(C$_3$,C$_3$) environments during high-temperature network rearrangement.}
    \label{fig:S10}
\end{figure}

Representative fragments in Figure~\ref{fig:S10}b show that transient \ce{C-C} bond breaking within an already developed carbon network, for example near a larger ring motif, can produce C$_2$(C$_3$,C$_3$) sites with a neighboring \ce{C-C} separation well beyond the bonding cutoff of \SI{1.85}{\angstrom}.
The lifetime analysis in Figure~\ref{fig:S10}a further shows that C$_2$(C$_3$,C$_3$) environments have much shorter mean, median, and maximum lifetimes than the other C$_2$ environments. 
Thus, C$_2$(C$_3$,C$_3$) species are best interpreted as transient high-energy intermediates associated with local bond rearrangement.

\clearpage

\setstretch{1.1}

\section{Supplementary references}
\vspace{3mm}

\providecommand{\latin}[1]{#1}
\makeatletter
\providecommand{\doi}
  {\begingroup\let\do\@makeother\dospecials
  \catcode`\{=1 \catcode`\}=2 \doi@aux}
\providecommand{\doi@aux}[1]{\endgroup\texttt{#1}}
\makeatother
\providecommand*\mcitethebibliography{\thebibliography}
\csname @ifundefined\endcsname{endmcitethebibliography}  {\let\endmcitethebibliography\endthebibliography}{}